\documentclass[aps, prd, reprint, letterpaper, preprintnumbers, floatfix, superscriptaddress]{revtex4-1} 

\usepackage{amsmath}
\usepackage{amssymb}
\usepackage[australian]{babel}
\usepackage{braket}
\usepackage{graphicx}
\usepackage{hyperref}
\usepackage{nicefrac}
\usepackage{pgf}
\usepackage{placeins}
\usepackage{siunitx}
\usepackage[caption=false]{subfig}
\usepackage[normalem]{ulem}
\usepackage{wasysym}
\usepackage{xcolor}

\usepackage{amsmath}
\usepackage{amssymb}
\usepackage{amsfonts}
\usepackage{amsthm}
\usepackage{bm}
\usepackage{xfrac}

\allowdisplaybreaks[1]
\newcommand{\conteqn}{\nonumber\\*}
\newcommand{\neweqn}{\\}

\newcommand{\ee}[0]{\mathrm{e}}
\newcommand{\ii}[0]{i}

\newcommand{\identity}[0]{\mathbb{I}}

\newcommand{\vect}[1]{\bm{#1}}
\newcommand{\unitvect}[1]{\smash{\widehat{\vect{#1}}}\;\!\vphantom{#1}}

\newcommand{\clovercoeff}{C_{SW}}

\newcommand{\dipoleYIntercept}{G_0}
\newcommand{\dipoleScale}{\Lambda}

\newcommand{\spatiallength}{L_s}

\newcommand{\temporalsites}{N_t}

\newcommand{\current}{\mathcal{J}}

\newcommand{\proj}{\projector{\vect{p}}}

\newcommand{\projm}{\projector{-\vect{p}}}

\newcommand{\projpm}{\projector{\pm\vect{p}}}
\newcommand{\projpmp}{\projector{\pm\vect{p}'}}

\newcommand{\minmom}[0]{\left|\vect{q}_\text{min}\right|}

\newcommand{\spinhalf}[0]{spin-\sfrac{1}{2}}

\newcommand{\transform}[1]{\xrightarrow{#1}}

\newcommand{\adjoint}[1]{\smash{\overline{#1}}\vphantom{#1}}

\newcommand{\transpose}{^\top}

\DeclareMathOperator{\Tr}{Tr}

\DeclareMathOperator{\sign}{sign}

\newcommand{\definedby}{\equiv}

\newcommand{\backderivLDm}[0]{{\overleftarrow{\nabla}^{{\mu}}}}
\newcommand{\backderivLDr}[0]{{\overleftarrow{\nabla}^{{\rho}}}}
\newcommand{\basisLDm}[0]{{e^{{\mu}}}}
\newcommand{\basisLUm}[0]{{e^{{\mu}}}}

\newcommand{\deltaLDmLDn}[0]{{\delta^{{\mu}{\nu}}}}
\newcommand{\energySi}[1]{{E_{{\alpha}}(#1)}}
\newcommand{\epsilonCaCbCc}[0]{{\epsilon^{{a}{b}{c}}}}
\newcommand{\ffDipole}[1][Q^2]{{G_{{D}}(#1)}}
\newcommand{\ffDiracSi}[1][Q^2]{{F_{{1}\,{\alpha}}(#1)}}
\newcommand{\ffDirac}[1][Q^2]{{F_{{1}}(#1)}}
\newcommand{\ffElectricSi}[1][Q^2]{{G_{{E}\,{\alpha}}(#1)}}
\newcommand{\ffElectricSne}[1][Q^2]{{G_{{E}\,{n}}(#1)}}
\newcommand{\ffElectricSpr}[1][Q^2]{{G_{{E}\,{p}}(#1)}}
\newcommand{\ffElectric}[1][Q^2]{{G_{{E}}(#1)}}
\newcommand{\ffMagneticConv}[1][Q^2]{{G^{{\text{Conv.}}}_{{M}}(#1)}}
\newcommand{\ffMagneticPEVA}[1][Q^2]{{G^{{\text{PEVA}}}_{{M}}(#1)}}
\newcommand{\ffMagneticSi}[1][Q^2]{{G_{{M}\,{\alpha}}(#1)}}
\newcommand{\ffMagneticSne}[1][Q^2]{{G_{{M}\,{n}}(#1)}}
\newcommand{\ffMagneticSpr}[1][Q^2]{{G_{{M}\,{p}}(#1)}}
\newcommand{\ffMagnetic}[1][Q^2]{{G_{{M}}(#1)}}
\newcommand{\ffPauliSi}[1][Q^2]{{F_{{2}\,{\alpha}}(#1)}}
\newcommand{\ffPauli}[1][Q^2]{{F_{{2}}(#1)}}
\newcommand{\forwderivLDm}[0]{{\overrightarrow{\nabla}^{{\mu}}}}
\newcommand{\forwderivLDr}[0]{{\overrightarrow{\nabla}^{{\rho}}}}
\newcommand{\gammaLCk}[0]{{\gamma^{{k}}}}
\newcommand{\gammaLUfive}[0]{{\gamma^{{5}}}}
\newcommand{\gammaLUfour}[0]{{\gamma^{{4}}}}
\newcommand{\gammaLUm}[0]{{\gamma^{{\mu}}}}
\newcommand{\gevectleftSiOi}[0]{{v_{{\alpha}\,{i}}}}
\newcommand{\gevectrightSiOi}[0]{{u_{{\alpha}\,{i}}}}
\newcommand{\gevectrightSiOj}[0]{{u_{{\alpha}\,{j}}}}

\newcommand{\indSb}[0]{{B}}

\newcommand{\indSdbl}[0]{{u_{\indSpr}}}

\newcommand{\indSi}[0]{{\alpha}}

\newcommand{\indSne}[0]{{n}}

\newcommand{\indSpr}[0]{{p}}

\newcommand{\indSsing}[0]{{d_{\indSpr}}}

\newcommand{\indexLCk}[1]{{{#1}^{{k}}}}
\newcommand{\indexLDe}[1]{{{#1}^{{\eta}}}}
\newcommand{\indexLDl}[1]{{{#1}^{{\lambda}}}}
\newcommand{\indexLDm}[1]{{{#1}^{{\mu}}}}
\newcommand{\indexLDn}[1]{{{#1}^{{\nu}}}}
\newcommand{\indexLDr}[1]{{{#1}^{{\rho}}}}
\newcommand{\indexLDs}[1]{{{#1}^{{\sigma}}}}
\newcommand{\interpOi}[2]{{\chi_{{i}}(#2)}}
\newcommand{\interpPEVAOip}[2]{{\chi_{{#1}\,{i'}}(#2)}}
\newcommand{\interpPEVAOi}[2]{{\chi_{{#1}\,{i}}(#2)}}
\newcommand{\interpbarPEVAOi}[2]{{\adjoint{\chi}_{{#1}\,{i}}(#2)}}

\newcommand{\interpoptPEVASi}[2]{{\phi^{{}}_{{#1}\,{\alpha}}(#2)}}
\newcommand{\interpoptbarPEVASi}[2]{{\adjoint{\phi}^{{}}_{{#1}\,{\alpha}}(#2)}}
\newcommand{\interpoptcouplingPEVASi}[1]{{z^{{}}_{{#1}\,{\alpha}}}}
\newcommand{\interpoptcouplingbarPEVASi}[1]{{\adjoint{z}^{{}}_{{#1}\,{\alpha}}}}

\newcommand{\linkvarDagLDm}[0]{{U^{{\dagger}\,{\mu}}}}
\newcommand{\linkvarDagLUm}[0]{{U^{{\dagger}\,{\mu}}}}
\newcommand{\linkvarLDm}[0]{{U^{{\mu}}}}
\newcommand{\linkvarLUm}[0]{{U^{{\mu}}}}
\newcommand{\massPhysSpr}[0]{{m^{{\mathrm{{Phys}}}}_{{p}}}}
\newcommand{\massSb}[0]{{m_{{B}}}}
\newcommand{\massSi}[0]{{m_{{\alpha}}}}
\newcommand{\massSn}[0]{{m_{{N}}}}

\newcommand{\mmEff}[0]{{\mu_{{\mathrm{{Eff}}}}}}
\newcommand{\mmSne}[0]{{\mu_{{n}}}}
\newcommand{\mmSpr}[0]{{\mu_{{p}}}}
\newcommand{\nucleoninterpOne}[0]{{\chi_{{1}}}}
\newcommand{\nucleoninterpTwo}[0]{{\chi_{{2}}}}
\newcommand{\projectorLCk}[0]{{\Gamma^{{k}}}}
\newcommand{\projectorLUe}[0]{{\Gamma^{{\eta}}}}
\newcommand{\projectorLUfour}[0]{{\Gamma^{{4}}}}
\newcommand{\projectorLUn}[0]{{\Gamma^{{\nu}}}}
\newcommand{\projectorLUs}[0]{{\Gamma^{{\sigma}}}}
\newcommand{\projector}[2][]{{\Gamma^{{#1}}_{\!{#2}}}}

\newcommand{\quarka}[0]{{q}}

\newcommand{\quarkbara}[0]{{\adjoint{q}}}

\newcommand{\quarkdownCb}[0]{{d^{{b}}}}

\newcommand{\quarkupCa}[0]{{u^{{a}}}}
\newcommand{\quarkupCc}[0]{{u^{{c}}}}

\newcommand{\sigmaLUmLUn}[0]{{\sigma^{{\mu}{\nu}}}}
\newcommand{\sigmaLUmLUr}[0]{{\sigma^{{\mu}{\rho}}}}
\newcommand{\sigmaLUrLUm}[0]{{\sigma^{{\rho}{\mu}}}}
\newcommand{\spinorSb}[1][s]{{u_{{B}}{(p, #1)}}}
\newcommand{\spinorSi}[1][s]{{u_{{\alpha}}{(p, #1)}}}

\newcommand{\spinorbarSb}[1][s]{{\adjoint{u}_{{B}}{(p, #1)}}}
\newcommand{\spinorbarSi}[1][s]{{\adjoint{u}_{{\alpha}}{(p, #1)}}}

\newcommand{\spinorpSi}[1][s']{{u_{{\alpha}}{(p', #1)}}}
\newcommand{\spinorpbarSi}[1][s']{{\adjoint{u}_{{\alpha}}{(p', #1)}}}

\newcommand{\threecfSi}[6][]{{\mathcal{G}^{{3}}_{{#1}}(#2\,;#3\,,#4\,;#5\,,#6\,;\indSi{})}}

\newcommand{\threecfprojSi}[7][]{{G^{{3}}_{{#1}}(#7\,;#2\,;#3\,,#4\,;#5\,,#6\,;\indSi{})}}

\newcommand{\twocfSi}[2]{{\mathcal{G}(#1\,;#2\,;\indSi{})}}

\newcommand{\twocfprojPEVAOiOj}[2]{{G_{{i}{j}}(#1\,;#2)}}

\newcommand{\twocfprojPEVASi}[2]{{G(#1\,;#2\,;\indSi{})}}

\newcommand{\twocfprojPEVA}[2]{{G(#1\,;#2)}}

\newcommand{\vectorcurrentLUfour}[1][]{{j^{{4}}_{{#1}}}}
\newcommand{\vectorcurrentLUl}[1][]{{j^{{\lambda}}_{{#1}}}}
\newcommand{\vectorcurrentLUm}[1][]{{j^{{\mu}}_{{#1}}}}
\newcommand{\vectorcurrentLUr}[1][]{{j^{{\rho}}_{{#1}}}}

\allowdisplaybreaks[1]

\begin{document}

\preprint{
\vbox{
\hbox{ADP-18-26/T1074}
}}

\title{Opposite-Parity Contaminations in Lattice Nucleon Form Factors}
\date{26 February 2019}

\author{Finn~M.~Stokes}
\affiliation{Special Research Centre for the Subatomic Structure of
  Matter,\\Department of Physics, University of Adelaide, South
  Australia 5005, Australia}
\affiliation{J\"ulich Supercomputing Centre, Institute for Advanced Simulation,\\
  Forschungszentrum J\"ulich, J\"ulich D-52425, Germany}
\author{Waseem~Kamleh}
\author{Derek~B.~Leinweber}
\affiliation{Special Research Centre for the Subatomic Structure of
  Matter,\\Department of Physics, University of Adelaide, South
  Australia 5005, Australia}

\begin{abstract}
    The recently-introduced Parity Expanded Variational Analysis (PEVA)
    technique allows for the isolation of baryon eigenstates at finite momentum
    free from opposite-parity contamination. In this paper, we establish the
    formalism for computing form factors of spin-\nicefrac{1}{2} states using
    PEVA\@. Selecting the vector current, we compare the electromagnetic form
    factors of the ground state nucleon extracted via this technique to a
    conventional parity-projection approach.
    Our results show a statistically significant discrepancy between the PEVA
    and conventional analyses. This indicates that existing calculations of
    matrix elements of ground state baryons at finite momentum can be affected
    by systematic errors of \(\sim\SI{20}{\percent}\) at physical quark masses.
    The formalism introduced here provides an effective approach to removing
    these systematic errors.
\end{abstract}

\maketitle

\section{Introduction\label{sec:introduction}}
Lattice QCD investigations of baryon matrix elements form a rich and varied
field of study. In such investigations, it is necessary to isolate the state
of interest from the tower of energy levels observed on the lattice. For ground
states, this can be achieved through simple Euclidean time evolution, but it
is common to use more advanced techniques such as multi-exponential fits, the
summation method, or variational analysis in order to extract the signal of
interest from earlier time slices and avoid the noise present in the tails of
correlation functions. Regardless of the specific technique used, when working
with spin-\nicefrac{1}{2} baryons it is typical to perform a simple
zero-momentum parity projection to the parity sector of interest,
significantly reducing the number of possible contaminating states before even
beginning the analysis. This technique works perfectly for at-rest baryons,
completely removing opposite-parity contaminations, but for matrix elements
where at least one of the initial or final state is boosted to non-trivial
momentum this is no longer the case. Since eigenstates with non-zero momentum
are not eigenstates of parity, the parity sectors are no longer well defined
and a na{\"\i}ve parity projection allows opposite-parity states to re-enter
the correlation functions.

In Ref.~\cite{Menadue:2013kfi}, we introduced the Parity Expanded Variational
Analysis (PEVA) technique to address this issue, applying it to the extraction
of the nucleon spectrum from two point correlation functions. In this paper, we
extend the formalism to the computation of matrix elements from three-point
correlation functions, and apply it to the specific example of calculating
the Sachs electric and magnetic form
factors\index{form factors!Sachs} \(\ffElectric\) and \(\ffMagnetic\) of
the ground-state proton and neutron. The Sachs form factors describe the
response of a baryon to the vector current\index{vector current}. At low
\(Q^2\), these form factors give information about the large-scale
electromagnetic structure of the state, such as its charge
radius\index{charge radius} and magnetic moment\index{magnetic moment}; at
high \(Q^2\) they give information about the short-distance internal
structure of the state. These form factors can be determined experimentally
to high accuracy. Computing them in \textit{ab-initio} lattice QCD provides an
important confrontation of theory with experiment.

In addition, computing
these form factors on the lattice gives us important insight into the
underlying physics. For example, on the lattice it is possible to
separately compute the contributions to the form factors from connected diagrams
(as studied in this paper) and disconnected diagrams\index{disconnected loops},
giving insight into the role sea quarks play in the structure of the
proton and the neutron. One can also alter the
electric charges of the quarks, readily illustrating the quark-flavour
structure of the nucleon.

We probe the values of these form factors by creating an incoming nucleon on the lattice,
having it interact with a vector current with some momentum transfer \(\vect{q}\), and
then annihilating the outgoing nucleon with a fixed momentum \(\vect{p}'\). By momentum
conservation, the incoming state must have momentum \(\vect{p} = \vect{p}' - \vect{q}\).
Due to the way we include the vector current on the lattice, we only consider a small
number of fixed momentum transfers \(\vect{q}\). By varying the three-momenta of the outgoing
state and hence the incoming state, we gain access to the form factors at a range of
\begin{equation}
Q^2=\vect{q}^2 - {(\sqrt{\massSn^2+\vect{p}'^2} - \sqrt{\massSn^2+\vect{p}^2})}^2\,.
\end{equation}
In particular, these boosts provide access to values close to
\(Q^2=0\), well below \(|\vect{q}_{\min}|^2 = {(2 \pi / \spatiallength)}^2\),
without requiring the use of twisted boundary conditions. By accessing a range of values,
we gain insight into the \(Q^2\) dependence of the form factors, and can make a
comparison with various models and experiment. By studying the low-\(Q^2\)
dependence of the electric form factor, we can make an \textit{ab-initio} determination
of the charge radius\index{charge radius} of the proton. In
addition, we observe that when considering the contributions
from each quark flavour independently, \(\ffElectric\) and \(\ffMagnetic\)
have a similar \(Q^2\) dependence in the range considered. Hence, we can 
access the magnetic dipole moments\index{magnetic moment} of the
proton and neutron by taking ratios of the
quark-sector form factors.

\section{Parity Expanded Variational Analysis\label{sec:peva}}
The process of extracting a baryonic excited-state spectrum via the PEVA
technique is presented in full in Ref~\cite{Menadue:2013kfi}. We present here
a brief summary of this process to introduce the notation and concepts
necessary to describe the extension to the computation of baryonic matrix
elements.

We begin with a basis of \(n\) conventional spin-\nicefrac{1}{2} operators
\(\left\{\interpOi{}{x}\right\}\) that couple to the states of interest.
Adopting the Pauli representation of the gamma matrices, we introduce the PEVA
projector
\(\projpm \definedby \frac{1}{4} \left(\identity + {\gammaLUfour}\right)
\left(\identity \pm \ii {\gammaLUfive} {\gammaLCk} \indexLCk{\unitvect{p}}\right)\).
When acting on a spinor, \(\projpm\) projects to states of definite helicity.
The choice of sign corresponds to choosing the sign of the projected helicity.
For two point correlation functions, this choice of helicity is arbitrary, and
both formulations of the projector give the same results.

We construct a set of basis operators
\begin{subequations}
\begin{align}
    \interpPEVAOi{\pm\vect{p}}{x} &\definedby \projpm \, \interpOi{}{x}\,,\neweqn
    \interpPEVAOip{\pm\vect{p}}{x} &\definedby \pm \projpm \, \gammaLUfive{} \, \interpOi{}{x} \label{eqn:primedoperator} \,.
\end{align}
\end{subequations}
We note that we use a Euclidean metric \(\deltaLDmLDn{}\), and hence there is
no need to distinguish between contravariant and covariant indices.

As described in Ref.~\cite{Bowler:1997ej}, three-dimensional smearing of the
operators breaks Lorentz invariance, and as a result can alter their
transformation properties, introducing extra terms
into the operator coupling proportional to \(\gammaLUfour\). However, due to
the structure of the PEVA projector, \(\projpm \, \gammaLUfour = \projpm\). As
a result, the additional Dirac structure introduced is removed by the PEVA
projection. The only remaining effect is that due to the factor of \(\gammaLUfive\)
introduced for the primed operator, the primed and unprimed operators may have
different couplings to each state. In the absence of three-dimensional
smearing they differ only by kinematic factors. Hence the PEVA technique
effectively handles the nontrivial Dirac structure arising from three-dimensional
smearing, without the requirement for any special treatment.

We then seek an optimised set of operators \(\interpoptPEVASi{\pm\vect{p}}{x}\)
that each couple strongly to a single energy eigenstate \(\indSi\). These
optimised operators are constructed as linear combinations of the basis operators
\begin{subequations}
\begin{align}
    \interpoptPEVASi{\pm\vect{p}}{x} &\definedby
      \sum_{i} \gevectleftSiOi(\vect{p}) \, \interpPEVAOi{\pm\vect{p}}{x} \,,\neweqn
    \interpoptbarPEVASi{\pm\vect{p}}{x} &\definedby
      \sum_{i} \gevectrightSiOi(\vect{p}) \, \interpbarPEVAOi{\pm\vect{p}}{x}\,,
\end{align}
\end{subequations}
where the sum is over both the primed and unprimed operators. The coefficients
\(\gevectleftSiOi(\vect{p})\) and \(\gevectrightSiOi(\vect{p})\) can be found
as the left and right generalised
eigenvectors~\cite{Michael:1985ne,Luscher:1990ck} of
\(\twocfprojPEVA{\vect{p}}{t+\Delta{}t}\) and \(\twocfprojPEVA{\vect{p}}{t}\),
where the correlation matrix
\begin{align}
    \twocfprojPEVAOiOj{\vect{p}}{t}
    &\definedby
      \Tr\!\left(\!\sum_{\vect{x}} \ee^{-\ii \vect{p}\cdot\vect{x}}
      \braket{\Omega|\interpPEVAOi{\pm\vect{p}}{x}\,
        \interpbarPEVAOi{\pm\vect{p}}{0}|\Omega}\! \right),
\end{align}
with \(i\) and \(j\) ranging over both the primed and unprimed operators.

The choice of sign in
the PEVA projector has no effect on the values of these two-point correlation
functions. This follows from the Dirac structure of the baryons and the
standard trace properties of the associated gamma matrices. We note that
the basis operators occur twice in the
correlation function expression, for both creation and annihilation.
In the top-left block, both operators are unprimed and the contributions
from the cross-parity term in the projector (\(\pm \ii {\gammaLUfive} {\gammaLCk} \indexLCk{\unitvect{p}}\))
are zero, so the overall sign is consistent, regardless of the choice of \(\proj\)
and \(\projm\). In the bottom-right block, both operators are primed, so two factors
of \(\pm 1\) are introduced by the overall sign of \(\interpPEVAOip{\pm\vect{p}}{x}\)
in Eq.~\eqref{eqn:primedoperator},
and the contributions from the cross-parity term are once again zero, so the
factors of \(\pm 1\) cancel and the overall sign is consistent. In the
off-diagonal blocks, the contributions to the correlator are only from the
cross-parity term, so there is an overall factor of \(\pm 1\), which cancels with
the factor of \(\pm 1\) from the single primed operator in each of these blocks.
Hence, the values of the two-point correlation functions will be the same
regardless of the choice of \(\proj\) or \(\projm\). As a
result, the coefficients for constructing \(\interpoptPEVASi{+\vect{p}}{x}\)
and \(\interpoptPEVASi{-\vect{p}}{x}\) are identical, up to a choice of overall
sign of the eigenvector. We choose the eigenvector sign to ensure the operators
match at zero momentum, where the choice of \(\pm\vect{p}\) has no effect on
the physics.

Using the optimised operators, we can construct the eigenstate-projected
two-point correlation function
\begin{align}
    &\twocfprojPEVASi{\vect{p}}{t} \conteqn
    &\qquad\definedby
      \Tr\left(\sum_{\vect{x}} \ee^{-\ii \vect{p}\cdot\vect{x}}
      \braket{\Omega|\,\interpoptPEVASi{\pm\vect{p}}{x}\,
        \interpoptbarPEVASi{\pm\vect{p}}{0}\,|\Omega} \right) \conteqn
    &\qquad=\gevectleftSiOi(\vect{p})\,
      \twocfprojPEVAOiOj{\vect{p}}{t}\, \gevectrightSiOj(\vect{p})\,.
\end{align}

\section{Baryon matrix elements\label{sec:matrixelements}}
\subsection{General matrix elements\label{sec:matrixelements:general}}
In this section, we establish the formalism to extend the PEVA technique to the
computation of baryon form factors. 
To perform the extension, we consider three-point correlation
functions\index{correlation function!three point}, inspecting their Dirac structure
to extract the signal of interest. We then take ratios with two point functions
to remove the time dependence and cancel out dependence on the interpolator
couplings. The calculations are performed in the most general kinematics
that can be realised.

Due to a lattice Ward identity associated with the conserved current,
the three-point correlation functions for the electric form factor normalised
to unit charge must approach the two-point correlation
functions exactly on a configuration-by-configuration basis as
\(Q^2 \transform{} 0\). As a result, the two- and three-point correlation
functions are highly correlated at low \(Q^2\). The ratios
we take facilitate the cancellation of statistical fluctuations, significantly
reducing the statistical uncertainties in our extracted form factors.

\index{correlation function!three point|(} 
By performing a parity-expanded variational analysis as described in
Sec.~\ref{sec:peva}, we construct optimised operators
\(\interpoptPEVASi{\pm\vect{p}}{x}\) that couple to each state \(\indSi{}\).
We can use these operators to calculate the three point correlation
functions\index{correlation function!three point}
\begin{align}
    &\threecfSi[\pm]{\current\!}{\vect{p}'\!}{\vect{p}}{t_2}{t_1}\conteqn
    &\qquad\definedby \sum_{\vect{x}_2,\vect{x}_2} \ee^{-\ii \vect{p}'\cdot\vect{x}_2} \,
        \ee^{\ii (\vect{p}' - \vect{p})\cdot\vect{x}_1}\conteqn
    &\qquad\qquad\quad\times\braket{\Omega|\,\interpoptPEVASi{\pm\vect{p}'}{x_2}\,
        \current(x_1)\,\interpoptbarPEVASi{+\vect{p}}{0}\,|\Omega}\,,
\end{align}
where \(\current(x)\) is some current operator\index{current operator}, which is
inserted with a momentum transfer \(\vect{q} = \vect{p}' - \vect{p}\). The
consideration of \(\threecfSi[-]{\current\!}{\vect{p}'\!}{\vect{p}}{t_2}{t_1}\)
(where the sink operator uses the opposite PEVA projector sign convention to
the source operator) is required to optimise the extraction of the form factors
for general kinematics. We note that it is sufficient to consider this change
of projector for the sink operator alone, leaving the source operator as
\(\interpoptbarPEVASi{+\vect{p}}{0}\) in all cases considered.

By inserting the complete set of states
\begin{equation}
    \identity = \sum_{\indSb{},p,s} \ket{\indSb{}\,; p\,; s} \bra{\indSb{}\,; p\,; s}
\end{equation}
on either side of the
current, and noting the use of Euclidean time\index{Euclidean space-time} and fixed boundary conditions (or negligible
backward-running state contributions), we can rewrite this three point correlation
function as
\begin{align}
    &\threecfSi[\pm]{\current\!}{\vect{p}'\!}{\vect{p}}{t_2}{t_1}\conteqn
        &\qquad= \sum_{s',s} \ee^{-\energySi{\vect{p}}\,t_1} \, \ee^{-\energySi{\vect{p}'}\,(t_2 - t_1)}\conteqn
        &\qquad \qquad\quad \times \braket{\Omega | \,\interpoptPEVASi{\pm\vect{p}'}{0}\, | \indSi{}\,; p'\,; s'}\conteqn
        &\qquad \qquad\quad \times \braket{\indSi{}\,; p'\,; s' | \,\current(0)\, | \indSi{}\,; p\,; s}\conteqn
        &\qquad \qquad\quad \times \braket{\indSi{}\,; p\,; s | \, \interpoptbarPEVASi{\vect{p}}{0}\, | \Omega}\,.
\end{align}
Note, the formalism presented here assumes perfect state isolation such that
each optimised operator couples only to a single state.

We see that the time dependence of this three point correlator is entirely contained
within exponentials of the energy, and the remaining structure depends on both the
overlap of the optimised operator with its corresponding state
\begin{equation}
    \braket{\Omega | \,\interpoptPEVASi{\pm\vect{p}}{0}\, | \indSi{}\,; p\,; s}
    = \interpoptcouplingPEVASi{\vect{p}}\,
        \sqrt{\frac{\massSi}{\energySi{\vect{p}}}} \; \projpm \, \spinorSi{}\,,\label{eqn:formfactors:interpoptcoupling}
\end{equation}
and the matrix element for the current operator,
\(\braket{\indSi\,; p'\,; s' | \,\current(0)\, | \indSi\,; p\,; s}\).

As we will see below, this Euclidean time dependence, along with the scalar
factors relating to the coupling between the optimised operator and the state
\(\indSi\) can be removed by taking appropriate ratios with the two point
correlation functions
\begin{align}
    \twocfprojPEVASi{\vect{p}}{t} \qquad\qquad\qquad& \conteqn
    = \Tr\Big(\sum_{s} \ee^{-\energySi{\vect{p}}\,t}
    & \braket{\Omega|\,\interpoptPEVASi{\pm\vect{p}}{0}\, | \indSi{}\,; p\,; s}\conteqn
    \times & \braket{\indSi{}\,; p\,; s | \,\interpoptbarPEVASi{\pm\vect{p}}{0}\,|\Omega}\Big)\,.
\end{align}

\subsection{Vector current matrix elements\label{sec:matrixelements:vector}}
In this paper, we investigate the electromagnetic properties of the proton and
neutron by choosing the current operator \(\current(x)\) to be the vector
current\index{vector current}. In particular, we use the tree-level \(O(a)\)-improved~\cite{Martinelli:1990ny}
conserved vector current\index{vector current}
used in Ref.~\cite{Boinepalli:2006xd},
\begin{equation}
    \vectorcurrentLUm[CI]{}(x) \definedby \vectorcurrentLUm[C]{}(x)
    + \frac{r}{2}\,a\, \quarkbara{}(x)\left(\backderivLDr{} + \forwderivLDr{}\right)
            \sigmaLUrLUm{}\,\quarka{}(x)\,,
\end{equation}
where \(r\) is the Wilson parameter, and
\begin{subequations}
\begin{align}
    \forwderivLDm{} \quarka{}(x) \definedby
        \frac{1}{2}\big(&\linkvarLDm{}(x)\,\quarka{}(x+\basisLDm{})\conteqn
            &- \linkvarDagLDm{}(x-\basisLDm{})\,\quarka{}(x-\basisLDm{})\big)\neweqn
    \quarkbara{}(x) \backderivLDm{} \definedby
        \frac{1}{2}\big(&\quarkbara{}(x+\basisLDm{})\,\linkvarDagLDm{}(x)\conteqn
            &- \quarkbara{}(x-\basisLDm{})\,\linkvarLDm{}(x-\basisLDm{})\big)\,.
\end{align}
\end{subequations}
This current is derived from the standard conserved vector current for the
Wilson action, symmetrised around a lattice site
\begin{align}
    \vectorcurrentLUm[C]{}(x) \definedby \frac{1}{4}\Big[\,
        &\quarkbara{}(x) \,(\gammaLUm{} - r)\,
            \linkvarLUm{}(x)\,\quarka{}(x+\basisLUm{})\conteqn
        + \,&\quarkbara{}(x+\basisLUm{}) \,(\gammaLUm{} + r)\,
            \linkvarDagLUm{}(x)\,\quarka{}(x)\conteqn
        + \,&\quarkbara{}(x-\basisLUm{}) \,(\gammaLUm{} - r{})\,
            \linkvarLUm{}(x-\basisLUm{})\,\quarka{}(x)\conteqn
        + \,&\quarkbara{}(x) \,(\gammaLUm{} + r)\,
            \linkvarDagLUm{}(x-\basisLUm{})\,\quarka{}(x-\basisLUm{})\Big]\,.
\end{align}

As all terms in the improved conserved current are one-link terms,
the corrections to the tree-level couplings do not encounter large
non-perturbative mean-field improvement corrections associated
with tadpole contributions.

This choice of current operator gives the matrix element
\begin{align}
    &\braket{\indSi\,; p'\,; s' | \,\vectorcurrentLUm[CI]{}(0)\, | \indSi\,; p\,; s}\conteqn
    &\qquad = \sqrt{\frac{\massSi}{\energySi{\vect{p}}}} \,
              \sqrt{\frac{\massSi}{\energySi{\vect{p}'}}} \;
              \spinorpbarSi{} \conteqn
    &\qquad\qquad\quad\times \left(\gammaLUm{} \, \ffDiracSi
        - \frac{\sigmaLUmLUn{}\,\indexLDn{q}}{2 \massSi{}} \, \ffPauliSi\right)\conteqn
    &\qquad\qquad\quad\times\spinorSi{}\,,\label{eqn:formfactors:matrixelement}
\end{align}
where \(Q^2 = \vect{q}^2 - {\left(\energySi{\vect{p}'} - \energySi{\vect{p}}\right)}^2\)
is the squared four-momentum with the conventional sign, and the invariant scalar
functions \(\ffDirac\) and \(\ffPauli\) are respectively the
Dirac\index{form factors!Dirac} and Pauli\index{form factors!Pauli} form factors.

\begin{widetext}
Hence, using Eqs.~\eqref{eqn:formfactors:interpoptcoupling}
and~\eqref{eqn:formfactors:matrixelement} we can rewrite the correlator as
\begin{align}
    \threecfSi[\pm]{\vectorcurrentLUm[CI]{}}{\vect{p}'\!}{\vect{p}}{t_2}{t_1}
        & = \sum_{s',s} \ee^{-\energySi{\vect{p}}\,t_1} \, \ee^{-\energySi{\vect{p}'}\,(t_2 - t_1)}
        \frac{\massSi}{\energySi{\vect{p}}} \, \frac{\massSi}{\energySi{\vect{p}'}} \,
            \interpoptcouplingPEVASi{\vect{p}'\!} \, \interpoptcouplingbarPEVASi{\vect{p}} \conteqn
        &\qquad \times \projpmp \, \spinorpSi{} \, \spinorpbarSi{}
        \left( \gammaLUm{} \ffDiracSi
            - \frac{\sigmaLUmLUn{}\indexLDn{q}}{2 \massSi{}} \ffPauliSi \right)
        \spinorSi{} \, \spinorbarSi{} \, \proj \,.
\end{align}
Using the spin sum\index{spin sum}
\begin{align}
    \sum_{s} \spinorSb{} \, \spinorbarSb{} = \frac{-\ii\,\gamma\cdot p + \massSb{}}{2 \massSb{}}\,,
\end{align}
the three-point function is
\begin{align}
    \threecfSi[\pm]{\vectorcurrentLUm[CI]{}}{\vect{p}'\!}{\vect{p}}{t_2}{t_1}
        & = \ee^{-\energySi{\vect{p}}\,t_1} \, \ee^{-\energySi{\vect{p}'}\,(t_2 - t_1)} \,
            \interpoptcouplingPEVASi{\vect{p}'\!} \, \interpoptcouplingbarPEVASi{\vect{p}} \conteqn
        &\qquad \times \projpmp \frac{-\ii\,\gamma\cdot p' + \massSi{}}{2\energySi{\vect{p}'}}
        \left( \gammaLUm{} \ffDiracSi
              - \frac{\sigmaLUmLUn{}\,\indexLDn{q}}{2 \massSi{}} \ffPauliSi \right)
        \frac{-\ii\,\gamma\cdot p + \massSi{}}{2\energySi{\vect{p}}} \, \proj \,.
\end{align}%
\index{correlation function!three point|)}

To extract our desired signal from this spinor structure, we can take the spinor trace
with some spin-structure projector\index{spin-structure projector}
\(\projector{S}\). This trace is then called the
spinor-projected three-point correlation function\index{correlation function!three point}
\begin{align}
    \threecfprojSi[\pm]{\vectorcurrentLUm[CI]{}}{\vect{p}'\!}{\vect{p}}{t_2}{t_1}{\projector{S}}
        &\definedby \Tr\!\left(\,\projector{S}\,
        \threecfSi[\pm]{\vectorcurrentLUm[CI]{}}{\vect{p}'\!}{\vect{p}}{t_2}{t_1}\, \right) \conteqn
        &= \ee^{-\energySi{\vect{p}}\,t_1} \, \ee^{-\energySi{\vect{p}'}\,(t_2 - t_1)} \,
            \interpoptcouplingPEVASi{\vect{p}'\!} \, \interpoptcouplingbarPEVASi{\vect{p}} \conteqn
        &\quad \times \Bigg( \Tr\!\left(\projector{S} \, \projpmp \,
            \frac{-\ii\,\gamma\cdot p' + \massSi{}}{2\energySi{\vect{p}'}} \,
            \gammaLUm{} \,
            \frac{-\ii\,\gamma\cdot p + \massSi{}}{2\energySi{\vect{p}}} \, \proj \right)
            \ffDiracSi\conteqn
        &\qquad - \Tr\!\left(\projector{S} \, \projpmp \,
            \frac{-\ii\,\gamma\cdot p' + \massSi{}}{2\energySi{\vect{p}'}}\,
            \frac{\sigmaLUmLUn{}\,\indexLDn{q}}{2 \massSi{}} \,
            \frac{-\ii\,\gamma\cdot p + \massSi{}}{2\energySi{\vect{p}}} \, \proj \right)
            \ffPauliSi\Bigg)\,.\label{eqn:formfactors:spinorprojected}
\end{align}
If we consider the function
\begin{align}
    F'_{\pm}(\projector{S},\, \current) \definedby
        8 \energySi{\vect{p}} \energySi{\vect{p}'} \, \Tr\!\left(\projector{S} \, \projpmp \,
        \frac{-\ii\,\gamma\cdot p' + \massSi{}}{2\energySi{\vect{p}'}} \, \current \,
        \frac{-\ii\,\gamma\cdot p + \massSi{}}{2\energySi{\vect{p}}} \, \proj \right)\,,
\end{align}
where the prime on \(F'_{\pm}(\projector{S},\, \current)\) denotes the presence
of the PEVA projectors, then we can express Eq.~\eqref{eqn:formfactors:spinorprojected}
as
\begin{align}
    \threecfprojSi[\pm]{\vectorcurrentLUm[CI]{}}{\vect{p}'\!}{\vect{p}}{t_2}{t_1}{\projector{S}}
        & = \ee^{-\energySi{\vect{p}}\,t_1} \, \ee^{-\energySi{\vect{p}'}\,(t_2 - t_1)} \,
            \interpoptcouplingPEVASi{\vect{p}} \, \interpoptcouplingbarPEVASi{\vect{p}} \, \frac{1}{8 \energySi{\vect{p}} \energySi{\vect{p}'}} \conteqn
        & \quad \times \left( F'_{\pm}(\projector{S}, \gammaLUm{}) \, \ffDiracSi
            - \frac{\indexLDn{q}}{2 \massSi{}} \, F'_{\pm}(\projector{S}, \sigmaLUmLUn{}) \, \ffPauliSi \right) \,.
\end{align}

These spinor-projected correlation functions\index{correlation function!three point} have a
nontrivial time dependence, which can be removed by constructing the ratio~\cite{Leinweber:1990dv}\index{ratio}
\begin{align}\label{eqn:formfactors:ratio}
    R_{\pm}(\vect{p}', \vect{p}\,; \alpha\,; r, s)
    &\definedby \, \sqrt{\left|\frac{
        \indexLDm{r} \, \threecfprojSi[\pm]{\vectorcurrentLUm[CI]{}}{\vect{p}'\!}{\vect{p}}{t_2}{t_1}{\indexLDn{s} \, \projectorLUn{}} \;
        \indexLDr{r} \, \threecfprojSi[\pm]{\vectorcurrentLUr[CI]{}}{\vect{p}}{\vect{p}'\!}{t_2}{t_1}{\indexLDs{s} \, \projectorLUs{}}}{
        \twocfprojPEVASi{\vect{p}'}{t_2} \, \twocfprojPEVASi{\vect{p}}{t_2}}\right|} \conteqn
        & \quad \times \sign\!\left(
            \indexLDl{r} \, \threecfprojSi[\pm]{\vectorcurrentLUl[CI]{}}{\vect{p}'\!}{\vect{p}}{t_2}{t_1}{\indexLDe{s} \, \projectorLUe{}}\right) \,,
\end{align}\index{R@\(R_{\pm}(\vect{p}', \vect{p}\,; \alpha\,; r, s)\)|see {ratio}}%
where \(\projectorLUfour{} = (\identity + \gammaLUfour{}) / 2\) and
\(\projectorLCk{} = (\identity + \gammaLUfour{}) (\ii \, \gammaLUfive{} \, \gammaLCk{}) / 2\) form
the basis for the spin projectors we use, and \(\indexLDm{r}\) and \(\indexLDm{s}\) are
coefficients selected to determine the form factors. Care is taken in selecting
\(\indexLDm{r}\) and \(\indexLDm{s}\) to ensure that the relevant values of
\(F'_{\pm}(\projector{S},\, \current)\) remain purely real.
    
In addition, as the momentum transfer \(\vect{q} \transform{} \vect{0}\),
charge conservation requires that
the temporal component of the three point correlator for the conserved vector
current becomes exactly proportional to the two point correlator on each
gauge field configuration, that is
\begin{equation}
    \threecfprojSi[\pm]{\vectorcurrentLUfour[CI]{}}{\vect{p}}{\vect{p}\!}{t_2}{t_1}{\indexLDn{s} \, \projectorLUn{}}
    \propto \twocfSi{\vect{p}}{t_2}\,.
\end{equation}
Because of this, taking the ratio in Eq.~\eqref{eqn:formfactors:ratio}
facilitates the cancellation of statistical fluctuations in the two- and
three-point correlators, providing results with small statistical uncertainties,
at least in the case of \(\ffElectric{}\).

We can then define the reduced ratio\index{reduced ratio},
\begin{align}
    \adjoint{R}_{\pm}(\vect{p}', \vect{p}\,; \alpha\,; r, s) &\definedby
        \sqrt{\frac{2 \energySi{\vect{p}}}{\energySi{\vect{p}}+\massSi{}}} \,
        \sqrt{\frac{2 \energySi{\vect{p}'}}{\energySi{\vect{p}'}+\massSi{}}} \,
        R_{\pm}(\vect{p}', \vect{p}\,; \alpha\,; r, s) \,.
\end{align}\index{Rbar@\(\adjoint{R}_{\pm}(\vect{p}', \vect{p}\,; \alpha\,; r, s)\)|see {reduced ratio}}
Taking this reduced ratio and substituting in the expressions for the projected
correlation functions, we obtain
\begin{align}
    \adjoint{R}_{\pm}(\vect{p}', \vect{p}\,; \alpha\,; r, s) &=
        \frac{\indexLDm{r}\,\indexLDn{s}}{%
            16 \energySi{\vect{p}} \energySi{\vect{p}'}(\energySi{\vect{p}}+\massSi{})\,(\energySi{\vect{p}'}+\massSi{})
        }\conteqn
        &\quad\times\left( F'_{\pm}(\projectorLUn{}, \gammaLUm{}) \, \ffDiracSi
        - \frac{\indexLDr{q}}{2 \massSi{}} \, F'_{\pm}(\projectorLUn{}, \sigmaLUmLUr{}) \, \ffPauliSi
        \right) \,.
\end{align}
\end{widetext}

By investigating the \(\indexLDm{r}\) and \(\indexLDs{s}\) dependence of this
ratio, we find that the clearest signals are given by
\begin{subequations}\label{eqn:formfactors:selectedratios}
\begin{align}
    R^{T}_{\pm} &= \frac{2}{1 \pm\, \unitvect{p} \cdot \unitvect{p}'} \;
        \adjoint{R}_{\pm}\left(\vect{p}', \vect{p}\,; \alpha\,; (1, \vect{0}), (1, \vect{0})\right)
        \,, \neweqn
    R^{S}_{\mp} &= \frac{2}{1 \pm\, \unitvect{p} \cdot \unitvect{p}'} \;
        \adjoint{R}_{\mp}\left(\vect{p}', \vect{p}\,; \alpha\,; (0, \unitvect{r}),(0, \unitvect{s})\right)\,,
\end{align}
\end{subequations}
where \(\unitvect{s}\) is chosen such that
\(\vect{p} \cdot \unitvect{s} = 0 = \vect{p}' \cdot \unitvect{s}\), \(\unitvect{r}\) is
equal to \(\unitvect{q} \times \unitvect{s}\), and the sign \(\pm\) in
Eq.~\eqref{eqn:formfactors:selectedratios} is chosen such that
\(1 \pm \unitvect{p} \cdot \unitvect{p}'\) is maximised. This choice maximises
the signal in the lattice determination of the correlation function ratios.

We can then find the Sachs electric and magnetic form factors\index{form factors!Sachs},
\begin{subequations}
\begin{align}
    \ffElectricSi{} &\definedby \ffDiracSi{} - \frac{Q^2}{{\left(2\massSi{}\right)}^2} \, \ffPauliSi{}\,, \neweqn
    \ffMagneticSi{} &\definedby \ffDiracSi{} + \ffPauliSi{}\,,
\end{align}
\end{subequations}
through appropriate linear combinations of \(R^{T}_{\pm}\) and \(R^{S}_{\mp}\).
A similar procedure can be applied to extract the relevant form factors from any
current.

We have shown how the PEVA technique can be applied to the calculation of baryon form
factors for arbitrary kinematics. Doing so ensures that these form factors are free
from opposite parity contaminations, up to residual contaminations arising from
the use of a finite operator basis.

\section{Sachs Electric Form Factor\label{sec:ge}}
\begin{table*}[hbt]
    \caption{\label{tab:formfactors:ensembles}Details of the gauge field ensembles
        used in this analysis.
        For each ensemble we list both the pion mass given in Ref.~\cite{Aoki:2008sm},
        with the lattice spacing set by hadronic inputs, and our determination
        of the the squared pion mass with the lattice spacing listed in the
        table, which is set by the Sommer parameter with
        \(r_0 = \SI{0.4921 \pm 0.0064}{\femto\meter}\)~\cite{Aoki:2008sm}. 
        }
    \sisetup{%
    table-number-alignment = center,
    table-figures-integer = 1,
    table-figures-decimal = 0,
    table-figures-uncertainty = 0,
}
\begin{tabular*}{\textwidth}{%
@{\extracolsep{\fill}}
        S[table-figures-integer = 3, table-figures-decimal = 0, table-figures-uncertainty = 0]
        S[table-figures-integer = 1, table-figures-decimal = 4, table-figures-uncertainty = 2]
        S[table-figures-integer = 1, table-figures-decimal = 4, table-figures-uncertainty = 3]
        S[table-figures-integer = 3, table-figures-decimal = 0, table-figures-uncertainty = 0]
        S[table-figures-integer = 1, table-figures-decimal = 0, table-figures-uncertainty = 0]
}
    \toprule{} \vspace{-9pt}\\
    {PACS-CS \(m_{\pi} \, / \,\si{\mega\electronvolt}\)} & {\(a \, / \, \si{\femto\meter}\)} & {\(m_{\pi}^2 \, / \, \si{\giga\electronvolt}\)} & {\# conf.} & {\# src per conf.} \\
    \colrule{} \vspace{-9pt}\\
    702 & 0.1022 \pm 0.0015 & 0.3884 \pm 0.0113 & 399 & 1 \\
    570 & 0.1009 \pm 0.0015 & 0.2654 \pm 0.0081 & 397 & 1 \\
    411 & 0.0961 \pm 0.0013 & 0.1525 \pm 0.0043 & 449 & 2 \\
    296 & 0.0951 \pm 0.0013 & 0.0784 \pm 0.0025 & 400 & 2 \\
    156 & 0.0933 \pm 0.0013 & 0.0285 \pm 0.0012 & 197 & 4 \\
    \botrule{}
\end{tabular*}
\end{table*}

We now apply this technique to calculate the Sachs electric form factors of the
proton and the neutron. This gives us insight into the distribution of charge
within these states.

The results in this paper are calculated on the PACS-CS
\((2+1)\)-flavour full-QCD ensembles~\cite{Aoki:2008sm}, made available through the
ILDG~\cite{Beckett:2009cb}. These ensembles use a \(32^3 \times 64\) lattice, and
employ a renormalisation-group improved Iwasaki gauge action\index{action:improved} with
\(\beta = 1.90\) and non-perturbatively \(O(a)\)-improved Wilson quarks, with
\(\clovercoeff = 1.715\). We use five ensembles, with stated pion masses
from \(m_{\pi}=\SI{702}{\mega\electronvolt}\) to \(\SI{156}{\mega\electronvolt}\)~\cite{Aoki:2008sm},
and set the scale using the Sommer parameter with \(r_0 = \SI{0.4921 \pm 0.0064}{\femto\meter}\)~\cite{Aoki:2008sm}. 
More details of the individual ensembles are presented in Table~\ref{tab:formfactors:ensembles},
including the squared pion masses in the Sommer scale. In these finite volumes,
the momentum is quantised in units of \(\minmom{} \definedby \frac{2\pi}{32a}\).
When fitting
correlators, the \(\chi^2 / \mathrm{dof}\) is calculated with the full covariance
matrix, and the \(\chi^2\) values of all fits are consistent with an appropriate
\(\chi^2\) distribution, as determined by a one-sided Kolmogorov-Smirnov
test comparing the full set of all \(\chi^2\) values for each number of degrees of
freedom to the corresponding \(\chi^2\) distribution.

The relativistic components of the baryon spinor are suppressed by the inverse
of the baryon mass. Across the pion masses considered here, the nucleon mass
ranges from \SI{1.418(9)}{\giga\electronvolt} to \SI{0.993(15)}{\giga\electronvolt}. As such,
at the lighter pion masses, the relativistic components of the
baryon spinor will be enhanced by a factor of \(\sim \num{1.5}\).
As a result, the parity-mixing at finite momentum
will be increasingly problematic. However, at lighter pion masses, the gauge
noise is more significant, and can occlude the parity-mixing effects
if the statistics are insufficient.

For the variational analyses in this paper, we begin with an eight-interpolator
basis is formed from the conventional \spinhalf{} nucleon interpolators
\begin{align}
    \nucleoninterpOne = &\epsilonCaCbCc [{\quarkupCa}\transpose \, (C\gammaLUfive) \, \quarkdownCb] \quarkupCc\,,\ \text{and}\conteqn
    \nucleoninterpTwo = &\epsilonCaCbCc [{\quarkupCa}\transpose \, (C) \, \quarkdownCb] \gammaLUfive \quarkupCc\,,
\end{align}
with \num{16}, \num{35}, \num{100}, or \num{200} sweeps~\cite{Mahbub:2013ala}
of gauge-invariant
Gaussian smearing\index{gauge-invariant Gaussian smearing}~\cite{Gusken:1989qx}
with a smearing fraction of \(\alpha = 0.7\), applied at the quark source
and sinks in creating the propagators. For the PEVA analyses, this basis is
expanded to sixteen operators as described in Section~\ref{sec:peva}. Before
performing the Gaussian smearing, the gauge links to be used are smoothed by
applying four sweeps of three-dimensional isotropic stout-link
smearing~\cite{Morningstar:2003gk} with \(\rho=0.1\).

To extract the form factors, we fix the source at time slice
\(\temporalsites/4=\num{16}\). As we use fixed boundary conditions in the time
direction, this ensures that the source is sufficiently separated from the
boundary to minimise boundary effects. Utilising the sequential source
technique~\cite{Bernard:1985ss}, we invert through the current,
fixing the current insertion at time slice \num{21}. We choose time slice \num{21}
by inspecting the projected two-point correlation functions associated with each state and
observing that excited-state contaminations are suppressed by time slice
\num{21}. This is evaluated by fitting the effective mass\index{effective energy} in this region to a
single state ansatz verifying that the full covariance \(\chi^2 / \text{dof}\)
is satisfactory. Choosing the current insertion time in this way allows us to
use knowledge of the time slice where the nucleon dominates the two point
function to insert the current with the expectation that excited-state
systematic errors are contained within the statistical uncertainties. As the
correlation matrix is formed in the PEVA approach, opposite-parity
contaminations are suppressed at both the source and the sink.

While this technique is a useful guide to choosing a current insertion time,
it does not guarantee the elimination of excited state effects. Indeed, as our
results will show, excited-state effects can be much worse in three-point
functions than two-point functions.

In choosing the variational parameters \(t_0\) and \(\Delta t\)
(as defined in Ref.~\cite{Menadue:2013kfi}), we have
implemented the criteria described in Ref.~\cite{Blossier:2009kd} and compared
it with other choices of the variational parameters.  In the baryon sector one
is always facing challenges with the rapid onset of statistical uncertainties.
Moreover, as explored in Ref.~\cite{Mahbub:2013ala}, the condition number of
the correlation matrices deteriorates as one progresses in Euclidean time.
Through a careful investigation we found that commencing the variational
analysis one slice after the source provides significantly smaller
uncertainties in the projected correlators while still providing excellent
plateau behaviour in the effective mass or energy provided the second time is
two or three time slices later. Thus for this work we chose \(t_0 = 17\) and
\(\Delta t = 2\).

We then extract the form
factors as outlined in Section~\ref{sec:matrixelements} for every
possible sink time and once again look for a plateau consistent with a
single-state ansatz.

\begin{table*}
    \caption{\label{tab:formfactors:kinematics}Different kinematics used in our
        analysis to access a range of \(Q^2\) values. The \(Q^2\) value listed
        is for the ground-state nucleon at the lightest pion mass of
        \(m_{\pi} = \SI{156}{\mega\electronvolt}\). The statistical error listed
        for \(Q^2\) comes from both the determination of the mass of the state
        and the conversion to physical units.}
    \sisetup{%
    table-number-alignment = center,
    table-figures-integer = 1,
    table-figures-decimal = 0,
    table-figures-uncertainty = 0,
}
\begin{tabular*}{\textwidth}{%
@{\extracolsep{\fill}}
        c
        c
        c
        S[table-figures-integer = 1, table-figures-decimal = 4, table-figures-uncertainty = 3]
}
    \toprule{} \vspace{-9pt}\\
    {Source momentum \(\vect{p} \, / \minmom{}\)} & {Sink momentum \(\vect{p}' / \minmom{}\)} & {Momentum transfer \(\vect{q} \, / \minmom{}\)} & {\(Q^2 \, / \, \si{\giga\electronvolt^2}\)} \\
    \colrule{} \vspace{-9pt}\\
    \((2,0,0)\) & \((3,0,0)\) & \((1,0,0)\) & 0.0833 \pm 0.0027 \\
    \((2,0,1)\) & \((3,0,1)\) & \((1,0,0)\) & 0.0902 \pm 0.0028 \\
    \((1,0,0)\) & \((2,0,0)\) & \((1,0,0)\) & 0.1248 \pm 0.0037 \\
    \((1,0,1)\) & \((2,0,1)\) & \((1,0,0)\) & 0.1301 \pm 0.0038 \\
    \((0,0,0)\) & \((1,0,0)\) & \((1,0,0)\) & 0.1655 \pm 0.0048 \\
    \((0,0,1)\) & \((1,0,1)\) & \((1,0,0)\) & 0.1665 \pm 0.0048 \\
    \((2,0,0)\) & \((3,1,0)\) & \((1,1,0)\) & 0.2211 \pm 0.0066 \\
    \((1,0,0)\) & \((2,1,0)\) & \((1,1,0)\) & 0.2647 \pm 0.0078 \\
    \((0,0,0)\) & \((1,1,0)\) & \((1,1,0)\) & 0.3191 \pm 0.0092 \\
    \((0,-1,0)\) & \((1,0,0)\) & \((1,1,0)\) & 0.3449 \pm 0.0100 \\
    \((1,0,0)\) & \((3,0,0)\) & \((2,0,0)\) & 0.4228 \pm 0.0131 \\
    \((0,0,0)\) & \((2,0,0)\) & \((2,0,0)\) & 0.5989 \pm 0.0174 \\
    \((-1,0,0)\) & \((1,0,0)\) & \((2,0,0)\) & 0.6898 \pm 0.0199 \\
    \botrule{}
\end{tabular*}
\end{table*}

Performing the sequential source technique through the current requires us to
choose our current operators and momentum transfers at inversion time. However,
this allows us to vary the sink momentum, and by extension
the source momentum, as well as varying the form of the interpolation functions
at the sink. This gives us access to several states, as well as a range of values of
\begin{equation}
    Q^2 = \vect{q}^2 - {\left( \energySi{\vect{p}'} - \energySi{\vect{p}} \right)}^2\,.
\end{equation}
In particular, values of \(Q^2\) well below that encountered in the frames with
\(\vect{p}\), or \(\vect{p}' = (0,0,0)\) are accessed via kinematics such as
\(\vect{p} = \minmom{}(1,0,0)\), \(\vect{p}' = \minmom{}(2,0,0)\).
The main alternative approach to accessing \(Q^2\) values in
this region is to use twisted boundary conditions\index{boundary condition} to change the momentum discretisation,
allowing the valence quarks to take different momentum values from the sea quarks. In our approach, all
momenta attained by the valence quarks are present in the sea, and thus we avoid the
complexities of partial quenching effects inherent in the twisted boundary approach.
Table~\ref{tab:formfactors:kinematics} summarises the kinematics considered
herein.

To begin, we inspect the Euclidean time\index{Euclidean space-time} dependence of \(\ffElectric{}\),
extracted as outlined in Section~\ref{sec:matrixelements}.
We consider independently the connected contributions to \(\ffElectric{}\) from
single valence quarks of unit charge.
The two flavours considered are the doubly represented quark flavour,
or the up quark in the proton (\(\indSdbl{}\)); and the singly
represented quark flavour, or the down quark in the proton
(\(\indSsing{}\)).

In the case of perfect optimised operators, there should be no Euclidean time\index{Euclidean space-time}
dependence, and the extracted form factors should be perfectly constant (up
to statistical fluctuations) after the current insertion. However, in practice
a finite operator basis is insufficient to perfectly isolate each state, leading to residual
excited-state contaminations. These show up as enhanced or suppressed form
factors at early Euclidean times. In light of this, care must be taken to
select a Euclidean time region in which these excited-state contaminations
are suppressed and the single state ansatz is satisfied. To ensure this ansatz
is satisfied, we inspect the full covariance \(\chi^2 / \text{dof}\) of a
constant fit in the proposed plateau region, and require that it lies in an
acceptable range \(\lesssim 1.2\). At the same time, we ensure that we do not
fit excessively noisy points in the tail of the correlator, as they can serve
to suppress the \(\chi^2 / \text{dof}\) and hide the effects of excited state
contamination. In doing so, we sometimes find that the \(\chi^2 / \text{dof}\)
for a given plateau region differs significantly between the two analyses, and
this can lead to selected plateaus that start on different time slices.

\begin{figure}[tbp]
    {\centering
        \input{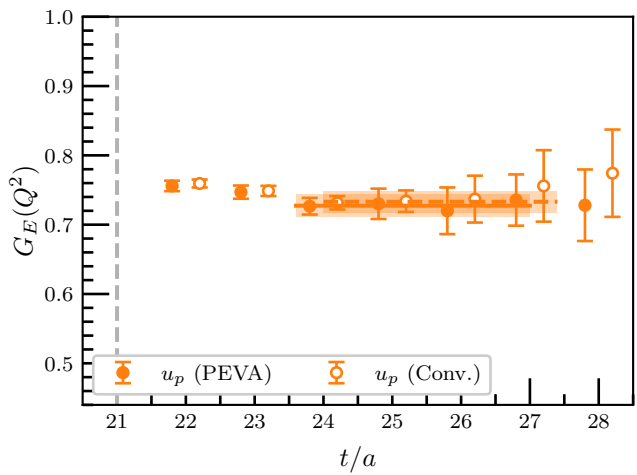}}
    \caption{\label{fig:groundstate:GEqs1:k5p000pp100}The contribution of
        the doubly represented quark flavour to the electric form factor of
        the ground-state nucleon at \(m_{\pi} = \SI{156}{\mega\electronvolt}\),
        for the lowest-momentum kinematics, providing
        \(Q^2 = \SI{0.1655 \pm 0.0048}{\giga\electronvolt^2}\). 
        Our fits to the plateaus are illustrated by shaded bands.
        We plot the conventional analysis with open markers and dashed fit lines
        and the new PEVA analysis with filled markers and solid fit lines.
        The source is at time slice 16, and the current is inserted at time
        slice 21. Both the conventional and PEVA fits are from time
        slice \num{24}--\num{27}.
        }
\end{figure}

\begin{figure}[tbp]
    {\centering
        \input{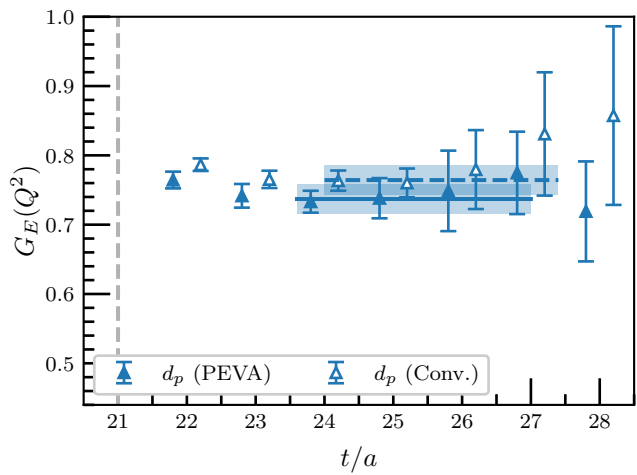}}
    \caption{\label{fig:groundstate:GEqs2:k5p000pp100}The contribution of the
        singly represented quark flavour to the electric form factor of
        the ground-state nucleon. The conventions used in this plot are the
        same as used in Fig.~\ref{fig:groundstate:GEqs1:k5p000pp100}.
        The kinematics are also the same, with \(m_{\pi} = \SI{156}{\mega\electronvolt}\),
        and \(Q^2 = \SI{0.1655 \pm 0.0048}{\giga\electronvolt^2}\). 
        Both the conventional and PEVA fits are from time
        slice \num{24}--\num{27}.}
\end{figure}

In Figs.~\ref{fig:groundstate:GEqs1:k5p000pp100} and~\ref{fig:groundstate:GEqs2:k5p000pp100}
we plot both PEVA and conventional extractions of \(\ffElectric\)
with respect to Euclidean sink time at the lightest
pion mass of \(m_\pi=\SI{156}{\mega\electronvolt}\) and the lowest-momentum
kinematics of \(\vect{p}=(0,0,0)\) and \(\vect{p}'=\minmom{}(1,0,0)\). We see that
starting from time slice \num{22}, which is immediately after the source, both
extractions of \(\ffElectric\) are quite flat across all time slices
considered. However, the errors on \(\ffElectric\) are sufficiently small to
identify a small Euclidean time dependence at early time slices. We find that
this dependence is suppressed by time slice \num{24} and are able to find a clear and
clean plateau from \num{24}--\num{27} for both extractions. For both quark
flavours considered, there is no significant
difference in the fit ranges, extracted values or errors between the two
extractions.

The conventional thought is that the opposite-parity
contaminations are small. Because they are from heavier states,
these contaminations are suppressed by  Euclidean time evolution.
We will see that this is not the case for the magnetic form factor, where we
will present direct evidence for important opposite parity contamination.
For the electric form factor, agreement between the PEVA and conventional
analyses can be maintained provided the opposite-parity contaminations
contribute to the form factor in a manner similar to that of the parity sector
under examination. As a result, despite
their continued presence, the opposite-parity contaminations do not
significantly perturb the value of the electric form factor.

\begin{figure}[tbp]
    {\centering
        \input{graphs_state0_fit_qs1_p000_pp100_k13727_GE.pgf}}
    \caption{\label{fig:groundstate:GEqs1:k2p000pp100}The contribution of the
        doubly represented quark flavour to the electric form factor of
        the ground-state nucleon at \(m_{\pi} = \SI{570}{\mega\electronvolt}\),
        for the lowest-momentum kinematics, providing
        \(Q^2 = \SI{0.1444 \pm 0.0044}{\giga\electronvolt^2}\). 
        The conventions used in this plot are the
        same as used in Fig.~\ref{fig:groundstate:GEqs1:k5p000pp100}.
        The PEVA fits start at time slice \num{26}, whereas the conventional
        fits start at time slice \num{28}. Note that this plot has been scaled
        up significantly to make the difference between the two fits visible.}
\end{figure}

At the heavier pion masses,
the statistical noise in the form factor extractions decreases, and the plateau
region shifts somewhat. However for all five masses, the qualitative behaviour
described above remains true, save for the following anomalies. At
\(m_\pi=\SI{570}{\mega\electronvolt}\), the plateaus from PEVA start two
time slices earlier than those from the conventional analysis. For example,
Fig.~\ref{fig:groundstate:GEqs1:k2p000pp100} shows the plateaus for the
doubly represented quark flavour. This is
potentially a signal of opposite-parity contaminations entering into the
analysis. However, there is no statistically significant difference in the
fit values from the two methods and the different plateaus do not show up at
any of the other masses considered, so it is inconclusive.

\begin{figure}[tbp]
    {\centering
        \input{graphs_state0_fit_qs1_p-100_pp100_k13781_GE.pgf}}
    \caption{\label{fig:groundstate:GEqs1:k5p-100pp100}Contributions of
        \(\indSdbl\) to the ground state \(\ffElectric\)
        at \(m_{\pi} = \SI{156}{\mega\electronvolt}\)
        for \(\vect{p} = \minmom{}(-1,0,0)\) and \(\vect{p}' = \minmom{}(1,0,0)\), providing
        \(Q^2 = \SI{0.690 \pm 0.020}{\giga\electronvolt^2}\). 
        The conventions used in this plot are the
        same as in Fig.~\ref{fig:groundstate:GEqs1:k5p000pp100}.
        The PEVA fit starts from time slice \num{23}, but the conventional
        analysis starts from time slice \num{24}.}
\end{figure}

\begin{figure}[tbp]
    {\centering
        \input{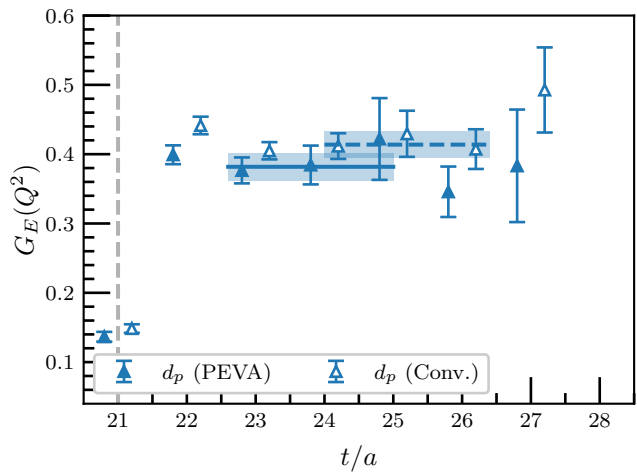}}
    \caption{\label{fig:groundstate:GEqs2:k5p-100pp100}Contributions of
        \(\indSsing\) to the ground state \(\ffElectric\).
        Pion mass and kinematics are as in
        Fig.~\ref{fig:groundstate:GEqs1:k5p-100pp100} above.
        The conventions used in this plot are the
        same as in Fig.~\ref{fig:groundstate:GEqs1:k5p000pp100}.
        The PEVA fit starts from time slice \num{23}, but the conventional
        analysis starts from time slice \num{24}.}
\end{figure}

We can also consider changing the momenta of the initial and final states, both
by changing the momentum transfer, and by boosting both the initial and final
states without changing the three-momentum transfer.

If we do this for the
\(m_{\pi} = \SI{156}{\mega\electronvolt}\) ensemble, where we previously found
consistent plateaus between PEVA and a conventional analysis, we find some
discrepancies. In Figs.~\ref{fig:groundstate:GEqs1:k5p-100pp100}
and~\ref{fig:groundstate:GEqs2:k5p-100pp100}, we boost the
initial state momentum to \(\vect{p}=\minmom{}(-1,0,0)\) and increase the momentum
transfer to \(\vect{q}=\minmom{}(2,0,0)\), leading to a significant increase in
\(Q^2\). In this case, we find that the PEVA plateaus start one time slice
earlier than the conventional plateaus. They have consistent plateau values,
but due to the earlier onset of the PEVA plateaus the statistical error is
reduced. These results suggest that there are contaminations in the extraction
of \(\ffElectric{}\) with the conventional analysis at
this mass. However the differences are not consistent across all higher-momentum
kinematics, and are not enough to categorically ascribe these problems to
opposite-parity contamination. We do note that when there is a difference in
the onset of the plateaus, the PEVA plateau always starts earlier.

For the other four masses, almost all kinematics have identical plateaus in
\(\ffElectric{}\) from both analyses, save for
\(m_{\pi} = \SI{570}{\mega\electronvolt}\), which once again has consistently
earlier plateaus for PEVA than the conventional analysis. It is unclear why
\(m_{\pi} = \SI{570}{\mega\electronvolt}\) has inconsistent plateaus at a range
of kinematics when the other three heavy masses don't. However, it is clear that
whatever opposite-parity contaminations are occurring, they are not
affecting the \(\ffElectric{}\) values extracted, at least within our current
statistical uncertainties.

Across all five masses, we
consistently find that at higher momenta there is more statistical noise in the
extraction of \(\ffElectric{}\). 

\begin{figure}[tbp]
    {\centering
        \input{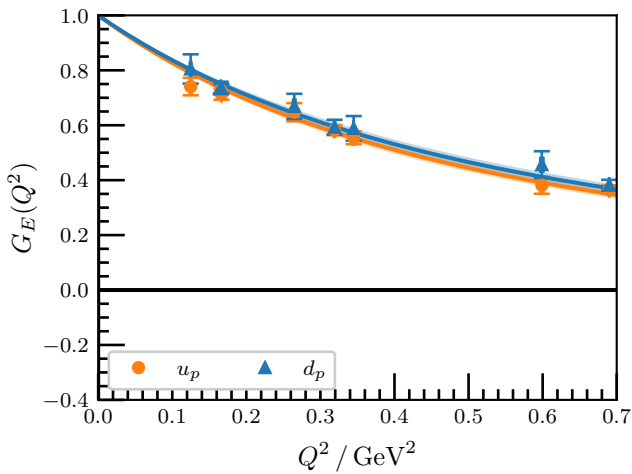}}
    \caption{\label{fig:groundstate:GE:k5Q2qs}Contributions from individual
        quark flavours to the electric form factor of the ground-state nucleon
        at \(m_{\pi} = \SI{156}{\mega\electronvolt}\). The shaded
        regions are dipole fits to the form factor, with lines indicating the
        central values. The \(y\)-axis intercept is fixed to one, as we are using
        an improved conserved vector current and the quarks are taken with
        unit charge.
        The errors on these fits are small enough that the
        shaded bands are barely distinguishable from the central lines.
        The fits correspond
        to a charge radius of \SI{0.684 \pm 0.019}{\femto\meter} for the doubly
        represented quark (\(\indSdbl\)) and \SI{0.659 \pm 0.021}{\femto\meter} for the
        singly represented quark (\(\indSsing\)).}
\end{figure}

In Fig.~\ref{fig:groundstate:GE:k5Q2qs}, we take the plateau values from each
of the kinematics listed in Table~\ref{tab:formfactors:kinematics} at
\(m_{\pi} = \SI{156}{\mega\electronvolt}\) and plot their
\(Q^2\) dependence. We exclude any kinematics for which we are unable to find a
clear plateau, or the variational analysis produces a negative generalised
eigenvalue (as negative eigenvalues indicate issues with the variational
analysis, and can cause problems with state identification). We see the
contributions from both quark flavours are very similar and each agrees well
with a dipole ansatz
\begin{equation}
    \ffDipole{} = \frac{\dipoleYIntercept{}}{{\left(1 + Q^2 / \dipoleScale{}^2\right)}^2}\,,
\end{equation}
with \(\dipoleYIntercept{}\) fixed to one, as we are working with single quarks of unit
charge. These fits correspond to an RMS charge radius\index{charge radius} of
\(\braket{r^2}^{\nicefrac{1}{2}} =
\sqrt{12}/\dipoleScale{} = \SI{0.684 \pm 0.019}{\femto\meter}\) for the doubly 
represented quark flavour and \SI{0.659 \pm 0.021}{\femto\meter} for the singly
represented quark flavour. That these values are smaller than the physical
expressions can be ascribed to the finite volume of the lattice~\cite{Hall:2012yx}.
For brevity, we omit similar plots for the other four masses.

\begin{figure}[tbp]
    {\centering
        \input{graphs_state0_nucleon_fit_k13781_GE.pgf}}
    \caption{\label{fig:groundstate:GE:k5Q2nucleon}\(\ffElectric\) for the
        ground-state proton and neutron
        at \(m_{\pi} = \SI{156}{\mega\electronvolt}\). The shaded
        region corresponds to a dipole fit to the proton form factor, with a charge
        radius of \SI{0.691 \pm 0.019}{\femto\meter}.}
\end{figure}

In order to compute the form factors of the proton,
\(\ffElectricSpr\), and neutron, \(\ffElectricSne\), we need to
take the correct linear combinations of the contributions from the doubly
and singly represented quark flavours to reintroduce the multiplicity of the
doubly represented quark and the physical charges of the up and down quarks.

\begin{table*}[hbt]
    \caption{\label{tab:groundstate:GE:k5Q2nucleon}\(\ffElectric\) at 
        \(m_{\pi} = \SI{156}{\mega\electronvolt}\) for all acceptable
        kinematics. We present results for the
        ground-state proton and neutron, as well as isovector combination
        \(\ffElectricSpr{} - \ffElectricSne{}\) (which
        is insensitive to disconnected loop corrections).}
    \sisetup{%
    table-number-alignment = center,
    table-figures-integer = 1,
    table-figures-decimal = 0,
    table-figures-uncertainty = 0,
}
\begin{tabular*}{\textwidth}{%
@{\extracolsep{\fill}}
        S[table-figures-integer = 1, table-figures-decimal = 4, table-figures-uncertainty = 3]
        S[table-figures-integer = 1, table-figures-decimal = 3, table-figures-uncertainty = 2]
        S[table-figures-integer = 1, table-figures-decimal = 3, table-figures-uncertainty = 2]
        S[table-figures-integer = 1, table-figures-decimal = 3, table-figures-uncertainty = 2]
}
    \toprule{} \vspace{-9pt}\\
    {\(Q^2 \, / \, \si{\giga\electronvolt^2}\)} & {\(\ffElectricSpr{}\)} & {\(\ffElectricSne{}\)} & {\(\ffElectricSpr{} - \ffElectricSne{}\)} \\
    \colrule{} \vspace{-9pt}\\
    0.1248 \pm 0.0037 & 0.719 \pm 0.034 & 0.043 \pm 0.028 & 0.677 \pm 0.051 \\
    0.1655 \pm 0.0048 & 0.724 \pm 0.018 & 0.006 \pm 0.010 & 0.718 \pm 0.023 \\
    0.1665 \pm 0.0048 & 0.705 \pm 0.025 & 0.014 \pm 0.015 & 0.691 \pm 0.038 \\
    0.2647 \pm 0.0078 & 0.640 \pm 0.034 & 0.014 \pm 0.020 & 0.626 \pm 0.044 \\
    0.3191 \pm 0.0092 & 0.578 \pm 0.017 & 0.007 \pm 0.009 & 0.571 \pm 0.018 \\
    0.3449 \pm 0.0100 & 0.543 \pm 0.024 & 0.023 \pm 0.023 & 0.520 \pm 0.038 \\
    0.5989 \pm 0.0174 & 0.357 \pm 0.029 & 0.050 \pm 0.020 & 0.307 \pm 0.035 \\
    0.6898 \pm 0.0199 & 0.360 \pm 0.017 & 0.011 \pm 0.011 & 0.350 \pm 0.025 \\
    \botrule{}
\end{tabular*}
\end{table*}

In Fig.~\ref{fig:groundstate:GE:k5Q2nucleon} and
Table~\ref{tab:groundstate:GE:k5Q2nucleon},
we present the electric form factors obtained by these combinations for
the lightest pion mass considered here.

In this work, we only consider connected contributions to the nucleon form
factors. There is no \textit{a priori} reason that the disconnected loops could
not be included in a PEVA calculation. They were simply omitted from the
analysis presented here for computational efficiency. The disconnected loop
contributions to the proton and neutron should be approximately the same
(exactly the same in our lattice calculations, as we are in the isospin
symmetric limit). Hence, if we take the isovector combination
\(\ffElectricSpr{} - \ffElectricSne{}\), the disconnected
loop\index{disconnected loops} contributions will cancel. The form factor
values for this combination are also presented in
Table~\ref{tab:groundstate:GE:k5Q2nucleon}.

The form factor for the neutrally charged neutron is close to zero for
all masses considered, as expected. Similar to the linear combinations taken for
the form factors, we can combine the squared charge radii\index{charge radius} from the individual
quark sectors with the appropriate multiplicities and charge factors to obtain
the squared charge radius\index{charge radius} of the neutron. For all five pion masses, we find a
small negative value. For example, at \(m_{\pi} = \SI{156}{\mega\electronvolt}\),
the neutron's squared charge radius is \SI{-0.022 \pm 0.009}{\femto\meter}.
This is qualitatively consistent with the negative squared charge radii observed
in experiment. A more quantitative discussion of this effect
requires knowledge of the disconnected loop contributions, which are not
considered in this work.

The form factor of the proton
matches well with a dipole fit with \(\dipoleYIntercept{}\) fixed to one (the
charge of the proton). As expected, the charge radii extracted from these
dipole fits approach the experimentally measured proton charge radius\index{charge radius}
from below as the pion mass is reduced towards the physical point.

As discussed in Refs.~\cite{Pohl:2013yb,Carlson:2015jba}, the exact physical
value of the proton radius\index{charge radius} has been a puzzle for the last seven years, since
precision laser spectroscopy of muonic hydrogen yielded a proton radius of
\SI{0.84087(39)}{\femto\meter}~\cite{Antognini:1900ns} in 2010. This value is
\SI{4.6}{\percent}, or \(5.6\sigma\) lower than the \textsc{codata} 2014
world average of \SI{0.8751(61)}{\femto\meter}~\cite{Mohr:2015ccw}, from a
combination of laser spectroscopy of electronic hydrogen and deuterium, and
elastic electron scattering. Recent precision results from new laser spectroscopy of
electronic hydrogen provide a proton radius of \SI{0.8335(95)}{\femto\meter}~\cite{Beyer79},
which agrees with the muonic hydrogen radius. This suggests that the discrepancy
is likely due to systematic errors in the existing results for electronic
hydrogen and elastic electron scattering.

\begin{figure}[tbp]
    {\centering
        \input{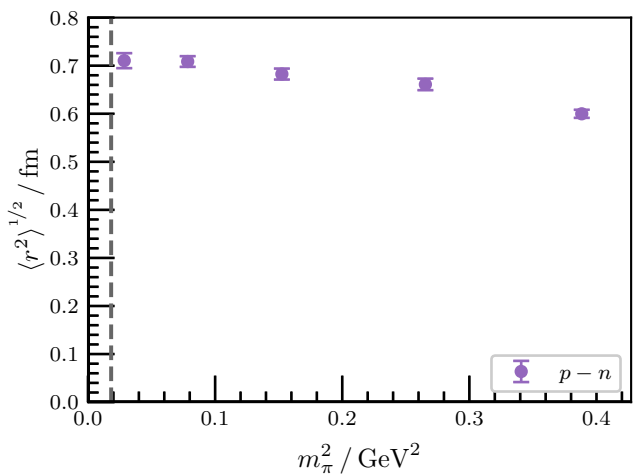}}
    \caption{\label{fig:groundstate:GE:proton}Quark-mass dependence of charge
    radii from dipole fits to the isovector combination
    \(\ffElectricSpr{} - \ffElectricSne{}\). We see a clear trend to larger
    radii as the pion mass approaches the physical point, represented by the
    dashed vertical line.}
\end{figure}

Returning to our results, in Fig.~\ref{fig:groundstate:GE:proton}, we
plot the charge radii obtained from dipole fits to the isovector combination
as a function of
the squared pion mass. We see that the pion-mass dependence is quite smooth,
suggesting that the structure of the state is fairly consistent at all five
masses considered here.
It has a clear trend of increasing charge radius as the mass is reduced.
This effect is in accord with the expectations of finite-volume chiral
perturbation theory~\cite{Hall:2012yx}.

For all pion masses and kinematics considered in this paper, in the specific
case of the electric form factor, there is no conclusive evidence of opposite
parity contaminations. Both the PEVA and conventional variational analysis show
clear and clean plateaus in \(\ffElectric{}\) with good excited state control.
This supports previous work demonstrating the utility of variational
analysis techniques in calculating baryon matrix elements~\cite{Owen:2012ts, Dragos:2016rtx}.
By using such techniques we are able to cleanly isolate precise values for the
Sachs electric form factor of the ground-state proton and
neutron.

\section{Sachs Magnetic Form Factor\label{sec:gm}}

\begin{figure}[tp]
    {\centering
        \input{graphs_state0_fit_p000_pp100_k13781_GM.pgf}}
    \caption{\label{fig:groundstate:GM:k5p000pp100}The contributions to the
        magnetic form factor from single quarks of unit charge for
        the ground-state nucleon at \(m_{\pi} = \SI{156}{\mega\electronvolt}\)
        for the lowest-momentum kinematics, providing
        \(Q^2 = \SI{0.1655 \pm 0.0048}{\giga\electronvolt^2}\). 
        We plot the conventional analysis with open markers and the new PEVA analysis
        with filled markers. Our fits to the plateaus are illustrated by shaded bands,
        with the central value indicated by dashed lines for the conventional analysis,
        and solid lines for the PEVA analysis.
        The plateau regions for both analyses are consistent, starting from time slice
        \num{23} for all four fits, but the value of the conventional plateau
        for the singly represented quark (\(\indSsing\)) has a magnitude
        approximately \SI{35}{\percent} lower than the PEVA plateau.}
\end{figure}

\begin{figure}[tp]
    {\centering
        \input{graphs_state0_fit_p000_pp100_k13727_GM.pgf}}
    \caption{\label{fig:groundstate:GM:k2p000pp100}The contributions to the
        magnetic form factor from single quarks of unit charge for
        the ground-state nucleon at \(m_{\pi} = \SI{570}{\mega\electronvolt}\)
        for the lowest-momentum kinematics, providing
        \(Q^2 = \SI{0.1444 \pm 0.0044}{\giga\electronvolt^2}\). 
        The plateau regions for both analyses are consistent, starting from time slice
        \num{23} for all four fits, but the value of the conventional plateau
        for the singly represented quark (\(\indSsing\)) has a magnitude
        slightly lower than the PEVA plateau.}
\end{figure}

Moving on to \(\ffMagnetic{}\), we once again begin with the lightest pion mass
and the lowest momenta. Here, we present results in terms of nuclear magnetons,
\(\mu_N \definedby \frac{e \hbar}{2 \massPhysSpr{}}\), defined in terms of the
physical proton mass, \(\massPhysSpr{}\). In
Fig.~\ref{fig:groundstate:GM:k5p000pp100}, we see that while the signal
is noisier than \(\ffElectric{}\), the excited-state contaminations present
at early Euclidean times are
less significant, and for both the PEVA and conventional analyses we are able
to find a plateau from time slice \num{23} to \num{25}. We are
cautious in fitting noisy data and restrict fit regimes to avoid large
fluctuations. Fig.~\ref{fig:groundstate:GM:k2p000pp100} illustrates a similar
plot for \(m_{\pi} = \SI{570}{\mega\electronvolt}\). Here the extended plateau
is from time slice \num{23} to \num{27} and is more representative of the three
heaviest pion masses considered.

Contrary to the electric case, there is a statistically significant difference
in the values of the plateaus from the PEVA and conventional analysis for the
singly represented quark flavour. If we take the correlated ratio of
\(\ffMagnetic{}\) from the conventional analysis to \(\ffMagnetic{}\) from the
PEVA analysis, we get a value of
\num{0.66 \pm 0.09}.
This ratio is clearly less than \num{1}, indicating that the magnitude of the
form factor is being significantly underestimated in the conventional analysis.
This suggests that despite finding a
plateau, the conventional analysis is being affected by opposite-parity
contaminations that are introducing a systematic error of approximately
\SI{35}{\percent}.

\begin{figure}[tbp]
    {\centering
        \input{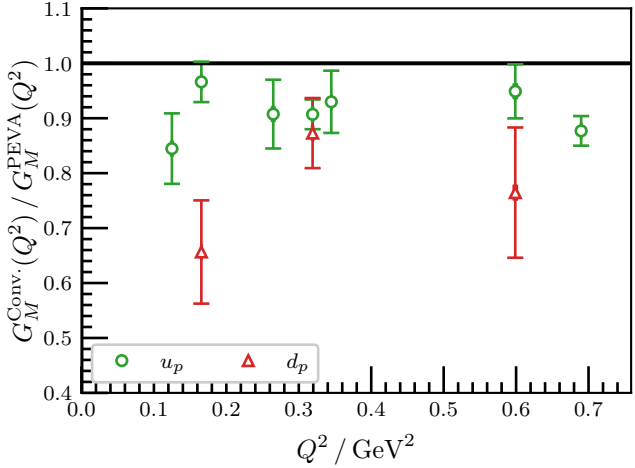}}
    \caption{\label{fig:groundstate:GM:k5ratio}Ratios of conventional plateaus
        to PEVA plateaus for ground state \(\ffMagnetic\) at
        \(m_{\pi} = \SI{156}{\mega\electronvolt}\). For clarity, ratios with
        large statistical errors have been excluded from the plot. If the
        plateaus were consistent, the points should be distributed
        about \num{1.0}. For the doubly represented
        quark flavour (\(\indSdbl\)), some kinematics match this expectation,
        but others sit more than one standard deviation low.
        The singly represented quark flavour (\(\indSsing\)) appears to show an
        even larger effect, with the ratios shifting even further away from
        unity albeit with larger statistical errors.}
\end{figure}

This is the strongest effect we see across all kinematics considered.
However, the conventional plateaus for other kinematics still show a statistically significant deviation
from the PEVA plateaus despite having the same fit regions and acceptable
\(\chi^2\) values. In Fig.~\ref{fig:groundstate:GM:k5ratio}, we plot the
correlated ratio discussed above for the kinematics that give the least noisy
extractions of \(\ffMagnetic{}\) with acceptable plateaus and
positive generalised eigenvalues.
We see that for the doubly represented quark sector, while some kinematics are
consistent with unity, others sit more than one standard deviation low.
The full covariance \(\chi^2 / \mathrm{dof}\) for an ansatz of unity across all
kinematics for which there are acceptable plateaus (including ratios not
on this graph) is \num{5.0}.
This indicates a significant disagreement between the two analyses, suggesting
that the conventional variational analysis is likely contaminated by
opposite-parity states. While it is not clear that the effect will be the same
across all kinematics, we can take a correlated weighted average across all of
the kinematics with valid plateaus to obtain an estimate for the size of the
effect. Doing so, we obtain a value of \num{0.92 \pm 0.02},
indicating that this quark flavour sees errors of \num{5}--\SI{10}{\percent}.
Removing the noisiest points as in Fig.~\ref{fig:groundstate:GM:k5ratio} does
not significantly alter these results, giving a \(\chi^2 / \mathrm{dof}\) of
\num{5.4} and
a weighted average of \num{0.91 \pm 0.02}.

The singly represented quark flavour (\(\indSsing\)) potentially shows an even larger
effect, with a weighted average of \num{0.83 \pm 0.05}.
The \(\chi^2 / \mathrm{dof}\) is lower due to larger statistical errors, taking
a value of \num{2.9}.
However, it is still quite large, indicating the difference between the two
analyses is significant. While removing the noisiest points as in
Fig.~\ref{fig:groundstate:GM:k5ratio} does increase the \(\chi^2 / \mathrm{dof}\)
to \num{6.4},
it does not significantly change the weighted average, giving
\num{0.78 \pm 0.05}.
These results indicate the presence of
opposite-parity contaminations, which introduce systematic errors of
\num{10}--\SI{20}{\percent} for \(\indSsing\), and perhaps more for some
specific kinematics.

As the states become less relativistic at larger quark masses, we see a
reduction in the
amount of parity mixing that occurs, and consequentially in the size of the
systematic errors, particularly at the heaviest two masses. However, we still
observe statistically significant deviations of the ratio below unity. For the
heaviest two masses of \SI{570}{\mega\electronvolt} and \SI{702}{\mega\electronvolt},
we see a systematic underestimation of the singly represented quark contributions
by \num{5}--\SI{10}{\percent} and at the
remaining masses of \SI{411}{\mega\electronvolt} and \SI{296}{\mega\electronvolt},
we see a \num{10}--\SI{15}{\percent} underestimation.

These results provide strong evidence for opposite-parity contaminations in
conventional extractions. These contaminations have a clear effect on the extracted
magnetic form factor at all five pion masses, on the order of \SI{10}{\percent}
for the doubly represented quark sector (\(\indSdbl\)) and \SI{20}{\percent}
for the singly represented quark sector (\(\indSsing\)) at the lighter masses.
Moving forward, use of the PEVA technique will be critical in
precision calculations of \(\ffMagnetic{}\) for the ground-state nucleon, for
which such systematic errors are unacceptable.

We now proceed to examine the extracted form factors.
In light of the opposite-parity contaminations present in the conventional
extractions, we focus only on the PEVA results.
Fig.~\ref{fig:groundstate:GM:k5Q2qs} shows the \(Q^2\) dependence of the
contribution to \(\ffMagnetic{}\) from each quark flavour at
\(m_\pi = \SI{156}{\mega\electronvolt}\). We see good agreement with a dipole
ansatz, with magnetic radii\index{magnetic radius} of \SI{0.514 \pm 0.030}{\femto\meter} for the
doubly represented quark flavour and \SI{0.85 \pm 0.11}{\femto\meter} for the
singly represented quark flavour.

\begin{figure}[tbp]
    {\centering
        \input{graphs_state0_qs_fit_k13781_GM.pgf}}
    \caption{\label{fig:groundstate:GM:k5Q2qs}Quark-flavour contributions to
        ground state \(\ffMagnetic\)
        at \(m_{\pi} = \SI{156}{\mega\electronvolt}\).
        The shaded regions are dipole fits to the form factor,
        corresponding to magnetic radii of \SI{0.514 \pm 0.030}{\femto\meter}
        for the doubly represented quark flavour (\(\indSdbl\)) and
        \SI{0.85 \pm 0.11}{\femto\meter} for the singly represented quark flavour
        (\(\indSsing\)).
        }
\end{figure}

\begin{figure}[tbp]
    {\centering
        \input{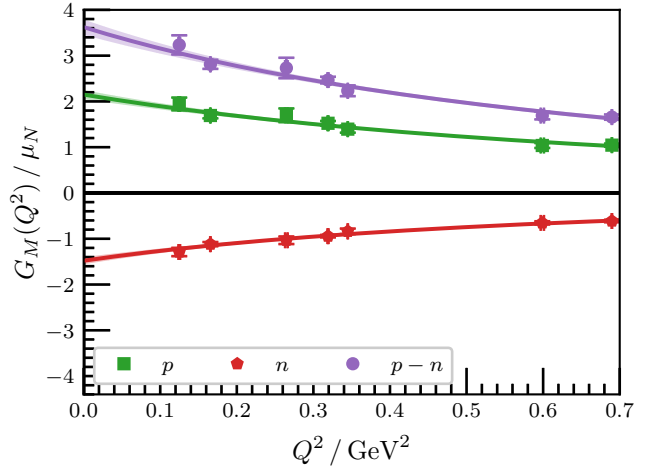}}
    \caption{\label{fig:groundstate:GM:k5Q2nucleon}\(\ffMagnetic\) for the
        ground-state proton and neutron
        at \(m_{\pi} = \SI{156}{\mega\electronvolt}\).
        The shaded region corresponds to a dipole fit to the form factor,
        with a magnetic radius of \SI{0.551 \pm 0.029}{\femto\meter} for the
        proton and \SI{0.618 \pm 0.031}{\femto\meter} for the neutron.
        We also include the isovector combination (\(\ffMagneticSpr{} - \ffMagneticSne{}\)),
        which is insensitive to disconnected loop corrections.}
\end{figure}

\begin{table*}[hbt]
    \caption{\label{tab:groundstate:GM:k5Q2nucleon}\(\ffMagnetic\) at 
        \(m_{\pi} = \SI{156}{\mega\electronvolt}\) for all acceptable
        kinematics. We present results for the
        ground-state proton and neutron, as well as isovector combination
        \(\ffMagneticSpr{} - \ffMagneticSne{}\) (which
        should be free from disconnected loop corrections).}
    \sisetup{%
    table-number-alignment = center,
    table-figures-integer = 1,
    table-figures-decimal = 0,
    table-figures-uncertainty = 0,
}
\begin{tabular*}{\textwidth}{%
@{\extracolsep{\fill}}
        S[table-figures-integer = 1, table-figures-decimal = 4, table-figures-uncertainty = 3]
        S[table-figures-integer = 1, table-figures-decimal = 2, table-figures-uncertainty = 2]
        S[table-sign-mantissa,table-figures-integer = 2, table-figures-decimal = 2, table-figures-uncertainty = 1]
        S[table-figures-integer = 1, table-figures-decimal = 2, table-figures-uncertainty = 2]
}
    \toprule{} \vspace{-9pt}\\
    {\(Q^2 \, / \, \si{\giga\electronvolt^2}\)} & {\(\ffMagneticSpr{} \, / \, \mu_N\)} & {\(\ffMagneticSne{} \, / \, \mu_N\)} & {\((\ffMagneticSpr{} - \ffMagneticSne{}) \, / \, \mu_N\)} \\
    \colrule{} \vspace{-9pt}\\
    0.1248 \pm 0.0037 & 1.94 \pm 0.14 & -1.29 \pm 0.09 & 3.23 \pm 0.21 \\
    0.1655 \pm 0.0048 & 1.69 \pm 0.06 & -1.12 \pm 0.05 & 2.81 \pm 0.10 \\
    0.2647 \pm 0.0078 & 1.70 \pm 0.15 & -1.03 \pm 0.08 & 2.73 \pm 0.22 \\
    0.3191 \pm 0.0092 & 1.52 \pm 0.05 & -0.94 \pm 0.03 & 2.47 \pm 0.07 \\
    0.3449 \pm 0.0100 & 1.39 \pm 0.08 & -0.84 \pm 0.06 & 2.23 \pm 0.12 \\
    0.5989 \pm 0.0174 & 1.04 \pm 0.05 & -0.65 \pm 0.04 & 1.69 \pm 0.08 \\
    0.6898 \pm 0.0199 & 1.05 \pm 0.03 & -0.61 \pm 0.03 & 1.66 \pm 0.06 \\
    \botrule{}
\end{tabular*}
\end{table*}

Similar to the electric form factor case described in Section~\ref{sec:ge},
we take linear combinations of the contributions from the doubly
and singly represented quark flavours to obtain the magnetic form factors of
the proton (\(\ffMagneticSpr\)) and neutron \(\ffMagneticSne\). In addition,
we can take the isovector combination (\(\ffMagneticSpr{} - \ffMagneticSne{}\))
to cancel out disconnected loop contributions.

We plot these combinations for the lightest pion mass in
Fig.~\ref{fig:groundstate:GM:k5Q2nucleon}, and present the values in
Table~\ref{tab:groundstate:GM:k5Q2nucleon}. At all five masses, the
magnetic form factors of both the proton and the neutron
agree well with a dipole fit. The magnetic radius\index{magnetic radius} obtained from each of
these fits is close to the electric charge radius of the proton extracted from
\(\ffElectric{}\) at the same pion mass.

\begin{figure}[tbp]
    {\centering
        \input{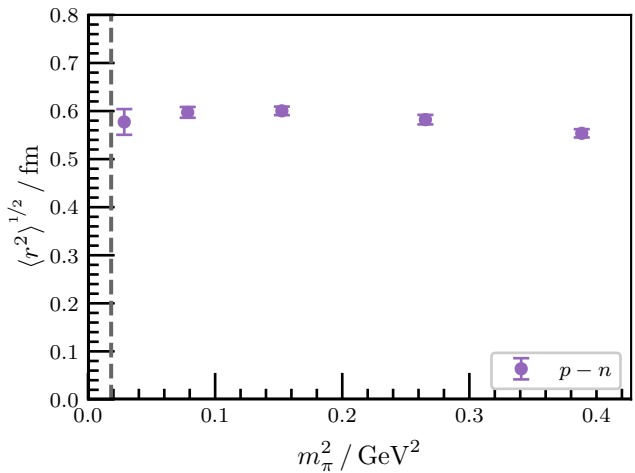}}
    \caption{\label{fig:groundstate:GM:nucleon}Quark-mass dependence of
        charge radii obtained from dipole fits to the isovector
        magnetic moment (\(\ffMagneticSpr{} - \ffMagneticSne{}\)).
        As in Fig.~\ref{fig:groundstate:GE:proton}, the dashed line
        corresponds to the physical pion mass.}
\end{figure}

In Fig.~\ref{fig:groundstate:GM:nucleon}, we plot the quark-mass dependence of
charge radii obtained from the dipole fits to the isovector combination
\(\ffMagneticSpr{} -\ffMagneticSne{}\). We can once again see a quark-mass
dependence, with increasing charge radius at lighter pion masses, aside from the
lightest mass, where the fit is getting too noisy to clearly distinguish a trend.
At the same time, \(\ffMagnetic[0]{}\) is increasing. This is in agreement with
expectations from chiral perturbation theory~\cite{Savage:2001dy, Leinweber:2002qb}.
\(\ffMagnetic[0]{}\) corresponds to the magnetic moment, which will be studied
in more detail in the next section.

In this section, we demonstrated the importance of the PEVA technique in
controlling systematic errors arising from opposite-parity contaminations
in extractions of the magnetic form factor for the ground-state nucleon. Due to
these opposite-parity contaminations, the conventional analysis underestimates
the contribution to the magnetic form factor from the singly represented quark
sector by \(\sim\SI{20}{\percent}\) at light pion masses, whereas the PEVA
technique removes the contaminations and gives improved results.

\section{Magnetic Dipole Moment\label{sec:mm}}%
\index{magnetic moment|(}
Returning to the individual quark flavour contributions, and noting that the
observed \(Q^2\) dependence of \(\ffElectric\) and \(\ffMagnetic\) is very
similar, we hypothesise that \(\ffMagnetic{}\) and
\(\ffElectric{}\) have the same scaling in \(Q^2\) over the range considered
here. If this hypothesis is valid, then the ratio of \(\ffMagnetic{}\) to
\(\ffElectric{}\) should be independent of \(Q^2\). Since we are working with
an improved conserved vector current, and single quarks of unit charge,
\(\ffElectric[0]{} = 1\) exactly, and \(\ffMagnetic[0]{}\) is the
contribution of the quark flavour
to the magnetic moment (up to scaling by the physical charge). Hence, the ratio
\begin{equation}
    \mmEff{}(Q^2) \definedby \frac{\ffMagnetic{}}{\ffElectric{}}\,,
\end{equation}
is expected to be constant in \(Q^2\), and equal to the contribution to the magnetic
moment from the given quark flavour.

Experimental results show that at high \(Q^2\), \(\mu \, \ffElectric / \ffMagnetic\)
diverges significantly from unity~\cite{Punjabi:2005wq}, so our hypothesis
must break down at sufficiently high \(Q^2\). However, over the low \(Q^2\) range
we consider here, these experimental results show that
\(\mu \, \ffElectric / \ffMagnetic\) is close to one, and hence within this
range \(\ffMagnetic / \ffElectric\) approximates the magnetic moment.

\begin{figure}[tbp]
    {\centering
        \input{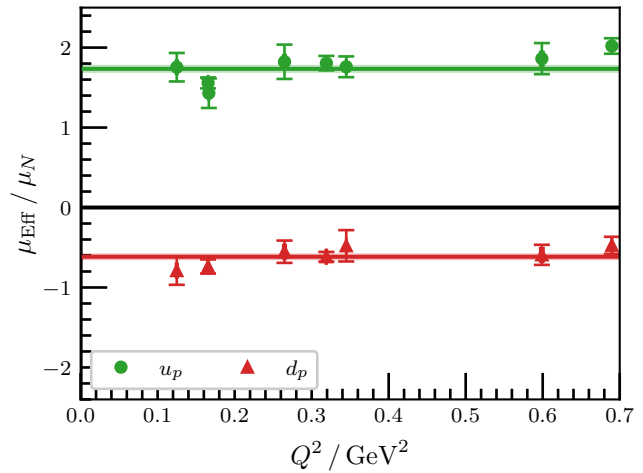}}
    \caption{\label{fig:groundstate:mm:k5}\(\mmEff{}(Q^2)\) for individual quark
        flavours in the ground state nucleon
        at \(m_{\pi} = \SI{156}{\mega\electronvolt}\). The narrow shaded bands are
        constant fits to the effective magnetic moment. 
        They correspond to magnetic moment contributions
        of \num{1.734 \pm 0.056} \(\mu_N\) for the doubly represented quark and 
        \num{-0.616 \pm 0.044} \(\mu_N\) for the singly represented quark.
        }
\end{figure}

\begin{figure}[tbp]
    {\centering
        \input{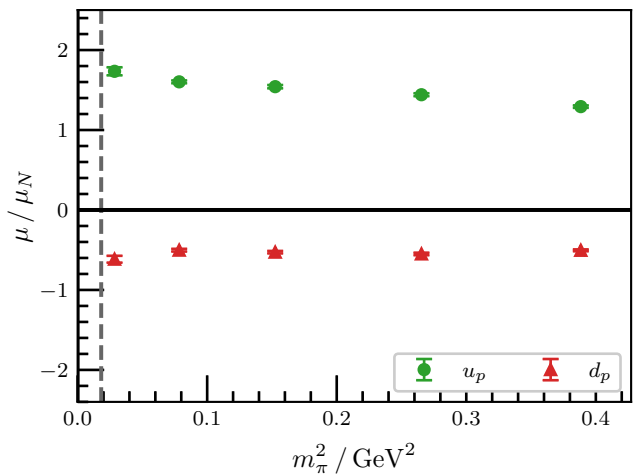}}
    \caption{\label{fig:groundstate:mm:qs}Pion-mass dependence of
        contributions to the ground-state magnetic moment from
        the doubly represented quark sector (\(\indSdbl\)) and
        the singly represented quark sector (\(\indSsing\)).
        The vertical dashed line shows the physical pion mass.}
\end{figure}

For all five pion masses, we find that \(\mmEff{}(Q^2)\) is indeed
approximately constant across the \(Q^2\) range considered. For example,
Fig.~\ref{fig:groundstate:mm:k5} shows the \(Q^2\)
dependence of \(\mmEff{}(Q^2)\) at the lightest pion mass. The remaining masses
show very similar \(Q^2\) dependence. By taking a constant
fit across all kinematics we obtain a estimate for the contributions to
the magnetic moment of the nucleon from single quarks of unit charge.
In the graphs shown here, the statistical errors on this fit are small, and
the shaded band showing these errors is almost indistinguishable from the solid
line indicating the central value of the fit. Fig.~\ref{fig:groundstate:mm:qs}
shows the pion mass dependence of these fits. These individual quark-flavour
contributions show a smooth pion-mass dependence with an enhancement of
the magnetic moments at low pion mass consistent with chiral perturbation
theory~\cite{Savage:2001dy, Leinweber:2002qb, Hall:2012pk}.

\begin{figure}[tbp]
    {\centering
        \input{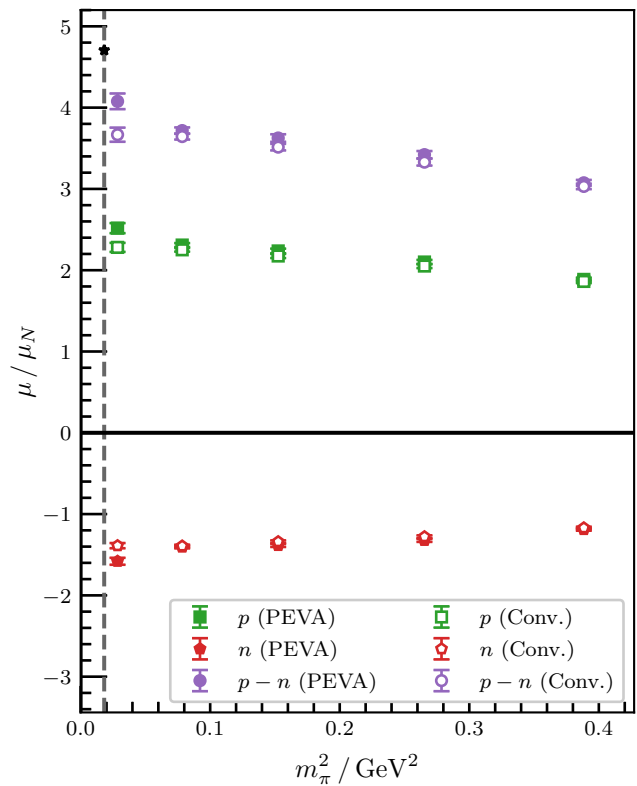}}
    \caption{\label{fig:groundstate:mm:nucleon}Pion-mass dependence of
        the extracted magnetic moment for the ground-state proton
        and neutron.
        To cancel out any disconnected loop contributions, we plot the isovector
        combination \(\mmSpr{} - \mmSne{}\). 
        As the physical point is approached, the trend in this combination
        approaches but doesn't quite reach the physical value of
        \(\num{4.70}\,\mu_N\)~\cite{Mohr:2015ccw}, represented by a black star.
        }
\end{figure}

\begin{table*}[hbt]
    \caption{\label{tab:groundstate:mm:nucleon}Magnetic moments at 
        \(m_{\pi} = \SI{156}{\mega\electronvolt}\) for all five pion masses.
        We present results for the
        ground-state proton and neutron, for both the PEVA analysis and the
        conventional parity projected analysis. We see that the conventional
        analysis consistently underestimates the magnetic moments, with the
        largest effect at smaller pion masses, where it reaches approximately
        \SI{10}{\percent}.}
    \sisetup{%
    table-number-alignment = center,
    table-figures-integer = 1,
    table-figures-decimal = 0,
    table-figures-uncertainty = 0,
}
\begin{tabular*}{\textwidth}{%
@{\extracolsep{\fill}}
        S[table-figures-integer = 1, table-figures-decimal = 4, table-figures-uncertainty = 3]
        S[table-figures-integer = 1, table-figures-decimal = 2, table-figures-uncertainty = 1]
        S[table-figures-integer = 1, table-figures-decimal = 2, table-figures-uncertainty = 1]
        S[table-sign-mantissa,table-figures-integer = 2, table-figures-decimal = 2, table-figures-uncertainty = 1]
        S[table-sign-mantissa,table-figures-integer = 2, table-figures-decimal = 2, table-figures-uncertainty = 1]
}
    \toprule{} \vspace{-9pt}\\
    {\(m_{\pi}^2 \, / \, \si{\giga\electronvolt^2}\)} & {\(\mmSpr{} \, / \, \mu_N\) (PEVA)} & {\(\mmSpr{} \, / \, \mu_N\) (Conv.)} & {\(\mmSne{} \, / \, \mu_N\) (PEVA)} & {\(\mmSne{} \, / \, \mu_N\) (Conv.)} \\
    \colrule{} \vspace{-9pt}\\
    0.3884 \pm 0.0113 & 1.89 \pm 0.02 & 1.86 \pm 0.02 & -1.19 \pm 0.01 & -1.17 \pm 0.01 \\
    0.2654 \pm 0.0081 & 2.10 \pm 0.03 & 2.05 \pm 0.03 & -1.32 \pm 0.02 & -1.28 \pm 0.02 \\
    0.1525 \pm 0.0043 & 2.24 \pm 0.03 & 2.17 \pm 0.03 & -1.39 \pm 0.02 & -1.34 \pm 0.02 \\
    0.0784 \pm 0.0025 & 2.31 \pm 0.02 & 2.25 \pm 0.02 & -1.41 \pm 0.02 & -1.39 \pm 0.02 \\
    0.0285 \pm 0.0012 & 2.52 \pm 0.06 & 2.28 \pm 0.06 & -1.58 \pm 0.04 & -1.39 \pm 0.03 \\
    \botrule{}
\end{tabular*}
\end{table*}

We can take the linear combinations discussed in Section~\ref{sec:ge}
to obtain the magnetic moments of the ground-state proton and
neutron. The quark mass dependence of these combinations is
illustrated in Fig.~\ref{fig:groundstate:mm:nucleon}, as well as in
Table~\ref{tab:groundstate:mm:nucleon}. We also present the equivalent magnetic
moment extractions from the conventional analysis. At the heavier pion masses,
the conventional analysis slightly underestimates the magnetic moment values.
At the lighter pion masses, this discrepancy grows rapidly, reaching approximately
\SI{10}{\percent} at the lightest pion mass considered here.

It is of interest to understand the origin of the difference between the PEVA and conventional
analyses at the lightest pion mass considered here. Inspecting the
excited state spectrum and the structure of the optimised operators at
the lightest two masses shows some difference, but no clear indication
of why the opposite-parity contaminations would be so much stronger at the
lightest mass. However, a detailed investigation of the negative parity
spectrum gives a hint at a possible cause. At the heavier pion masses, the
localised negative parity excitations have magnetic moments similar to quark
model expectations for the \(N^*(1535)\) and \(N^*(1650)\) resonances, as
presented in Ref.~\cite{Kamleh:2017lye}. However, at the lighter pion masses,
the magnetic moments shift away from the quark model expectations, suggesting
an increasing role for multi-particle states in the negative-parity
spectrum. This leads to a change in the nature of the
localised negative-parity states that couple well to the localised
operators used in this work, which in turn can significantly alter the effects
of opposite-parity contaminations on the ground state matrix elements.

The magnetic moments of the proton and neutron extracted by PEVA have a similar
quark mass dependence to the individual quark-flavour contributions and are close to the
experimental values of \num{2.7928473508(85)} \(\mu_N\) for the proton, and
\num{-1.913 042 73(45)} \(\mu_N\) for the neutron~\cite{Mohr:2015ccw}. The small discrepancy
between our results and the physical values is due to a combination of
disconnected loop\index{disconnected loops} contributions which are not included in our calculation, and
finite-volume\index{finite volume} effects. To avoid the disconnected loop
corrections, we compare the isovector combination \(\mmSpr{} - \mmSne{}\)
to the equivalent combination of the experimentally determined magnetic moments.
Doing this, we find that we underestimate the experimental value by around
\SI{10}{\percent}. This remaining discrepancy can be attributed to finite volume
corrections.

\begin{figure}[tbp]
    {\centering
        \input{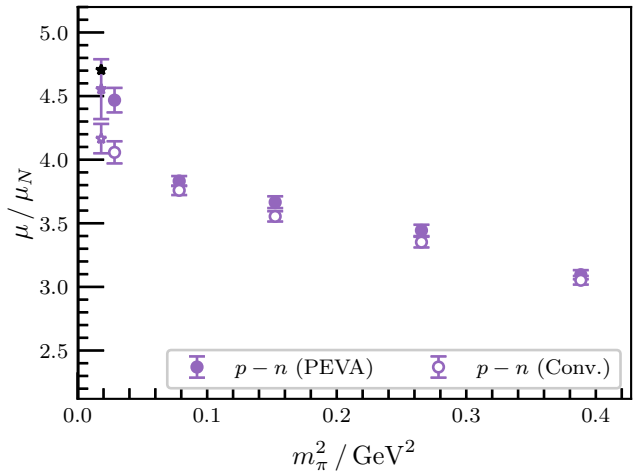}}
    \caption{\label{fig:groundstate:mm:isovector:iv}Pion-mass dependence of
        the extracted magnetic moment for the isovector combination
        with finite-volume corrections from Ref.~\cite{Hall:2012pk}.
        At the physical point, we present chiral extrapolations of the PEVA
        and conventional results with filled and open stars respectively.
        The PEVA result agrees well with the physical value of
        \(\num{4.70}\,\mu_N\)~\cite{Mohr:2015ccw}, represented by a black star.
        }
\end{figure}

To address these finite volume corrections, we consider the chiral effective
field theory study presented in Ref.~\cite{Hall:2012pk}.
Using this formalism, we estimate the finite volume corrections
to our our magnetic moment extractions at each pion mass. Our results
corrected to their predicted infinite-volume values are presented in
Fig.~\ref{fig:groundstate:mm:isovector:iv}. We see that this correction brings
our PEVA results much closer to the experimental value for the isovector
magnetic moment. Performing the full chiral extrapolation as in
Ref.~\cite{Hall:2012pk}, we find that the PEVA results are consistent with the
experimental result, while the conventional results sit many standard deviations
low. In this analysis, there is some instability in the extracted regulator
parameter, due to the smaller number of ensembles considered here. To account
for this, we have included in our analysis an additional systematic error not
considered in the reference, arising from varying \(m_{\pi,\max}^2\)
from \SI{0.39(1)}{\giga\electronvolt} to \SI{0.27(1)}{\giga\electronvolt}.
Because the PEVA results show a much stronger chiral curvature, the
extrapolation of these results is more sensitive to uncertainties in the
regulator, as seen in its larger error bars.

These results clearly indicate that the magnetic moment extracted
through the conventional analysis is significantly affected by opposite parity
contaminations, resulting in incorrect results. The PEVA analysis allows us to
remove these contaminations, bringing our results in line with experiment.%
\index{magnetic moment|)}

\section{Conclusion\label{sec:conclusion}}
In this paper, we extended the parity-expanded variational analysis (PEVA)
technique to the calculation of elastic
form factors, and applied it to calculating the Sachs electric and magnetic
form factors of
the ground-state proton and neutron. This required inspection of the Dirac
structure of the three point correlation function and careful selection of
appropriate spinor projectors to extract the desired form factors with maximised
signal.

Nucleon structure is a vibrant and rich field of study, and there have been
investigations of the Sachs electric and magnetic form factors of the
ground state nucleon spanning decades. In this paper we focused specifically
on the application of the PEVA technique~\cite{Menadue:2013kfi} to form factor
calculations and on the systematic errors introduced by opposite-parity
contaminations which may be present in conventional analyses.

We demonstrated the efficacy of variational analysis techniques in general, and
PEVA specifically, at controlling excited-state contaminations in the
electric form factor. Both the PEVA and conventional variational analysis show
clear and clean plateaus, supporting previous work demonstrating the utility of
variational analysis in calculating baryon matrix
elements~\cite{Dragos:2016rtx, Owen:2012ts}.

In the particular case of the magnetic form factor, we found evidence that the
conventional analysis was contaminated by opposite-parity states. For
the kinematics considered here, we observe \(\sim\SI{20}{\percent}\)
underestimation of the magnitude of the contributions to the magnetic form
factor from the singly represented quark flavour at the lighter pion masses.
The only difference in the interpolators is that opposite-parity contaminations
can be addressed in our new PEVA approach. The difference indicates these
contaminations are present in the standard variational approach. As the PEVA
approach provides additional interpolator degrees of freedom to improve the
ground state interpolating field, this is the improved interpolating field.

Further evidence of the improvement afforded by the PEVA approach is presented
in Ref.~\cite{Kamleh:2017lye}, where we explored excited states of the nucleon
at larger quark masses.  Only the PEVA approach is able to resolve magnetic
form factors in accord with constituent quark models. Quark models are renowned
for capturing the qualitative features of baryon magnetic moments at larger
quark masses.

These results indicate that existing calculations that do not take into account
opposite-parity contaminations may be affected by systematic errors on the
order of \SI{20}{\percent} at physical quark masses. As such, the PEVA
technique is critical for precision measurements of the nucleon form factors.

By utilising the PEVA technique and boosted-frame techniques, we are able to
successfully extract the Sachs
form factors of the ground-state nucleon at a range of \(Q^2\) values. These
extractions allow us to investigate the \(Q^2\) dependence of these form
factors. By taking ratios of the form factors, we are also able to extract
the magnetic moments of both the ground-state proton and
the ground-state neutron.

This paper has established the groundwork for applying the PEVA technique to
calculating baryonic matrix elements. The applications for future research
are broad. The techniques presented here could be applied directly to
the examination of other nucleon observables or excited state observables.
One possibility is to examine a matrix element that should vanish in the
ground-state nucleon, where any nonzero value is evidence of excited state
contamination. An extension is currently underway for the calculation of nucleon
transition form factors.

A straightforward application of the calculations performed here, with extra
statistics and a range of lattice spacings and volumes to quantify systematic
errors, could provide state-of-the-art ab-initio determinations of the nucleon
electromagnetic form factors. Such a study should also be able to fully quantify
the difference between PEVA and conventional results at the physical point
and confirm the disagreement between the conventional magnetic moment
extractions and the physical results.

Our results indicate that excited-state effects can be much worse in
three-point functions than two-point functions. The contrast between the good
agreement between PEVA and conventional extractions of the ground state mass
and the disagreement between extractions of the magnetic form factor indicate
that a plateau in the two-point correlator is thus insufficient evidence to be
confident that the three-point correlator will be free of excited-state
contaminations. Future calculations can consider a range of current insertion
times following the onset of ground-state dominance in the two-point function
to quantitatively assess excited-state systematics. 

An approach to applying PEVA to spin-\nicefrac{3}{2} states is under
consideration, allowing the issue of parity mixing to be addressed for a
wider range of baryonic states. Such an extension would also allow for the
inclusion of spin-\nicefrac{3}{2} states in the variational analysis for the
nucleon, addressing systematic errors that may arise from the mixing of
eigenstates of total angular momentum in moving frames. For a fully
comprehensive study of excited state contaminations of the ground state nucleon
multi-particle states should also be considered. Rigorously treating such states
requires non-local interpolators and this would be a challenging but worthwhile
avenue for further study, as it could simultaneously give additional insight
into the nature of nucleon resonances. A comprehensive study addressing all
of these sources of excited state contaminations could look at how strong the
effects of each contaminant are and determine which states are most responsible
for errors in the matrix elements extracted by conventional techniques.

A particularly interesting development in the field that could benefit greatly
from the application of the PEVA technique is the computation of non-forward
matrix elements from two point correlation functions via the Feynman-Hellmann
theorem~\cite{Chambers:2017tuf}.

\appendix*

\begin{acknowledgments}
This research was undertaken with the assistance of resources from the
Phoenix HPC service at the University of Adelaide, the National
Computational Infrastructure (NCI), which is supported by the Australian
Government, and by resources provided by the Pawsey Supercomputing Centre
with funding from the Australian Government and the Government of Western
Australia. These resources were provided through the National Computational
Merit Allocation Scheme and the University of Adelaide partner share. This
research is supported by the Australian Research Council through grants
no.\ DP140103067, DP150103164, and LE160100051.
\end{acknowledgments}

\bibliography{reference}

\end{document}